\documentstyle[times,psfig,epsf,12pt]{article}
\begin{document}
\newcommand{\ra}{\rightarrow}
\begin{flushright}
hep-ph/0410340
\end{flushright}
\smallskip
\begin{center}
{\Large \bf 
Summary of the Activities of the Working Group I \\ on High Energy and
Collider Physics}
\end{center}

\bigskip

\noindent{\bf Naba K. Mondal, Saurabh D. Rindani,
Pankaj Agrawal, Kaustubh Agashe, 
B. Ananthanarayan, Ketevi Assamagan, Alfred Bartl, Subhendu Chakrabarti, 
Utpal Chattopadhyay, Debajyoti Choudhury, 
Eung-Jin Chun, Prasanta K. Das, Siba P. Das, Amitava Datta,
Sukanta Dutta, Jeff Forshaw, Dilip K. Ghosh, 
Rohini M. Godbole, Monoranjan Guchait, Partha Konar, Manas Maity, 
Kajari Mazumdar,
Biswarup Mukhopadhyaya, Meenakshi Narain,
Santosh K. Rai, Sreerup Raychaudhuri, D.P. Roy, 
Seema Sharma, Ritesh K. Singh}

\bigskip
\centerline{\bf Abstract}
\begin{quote}
This is a summary of the projects undertaken by the Working Group I on
High Energy Collider Physics at the Eighth Workshop on High Energy Physics 
Phenomenology (WHEPP8) held at the Indian Institute of Technology, Mumbai, 
January 5-16, 2004.  The topics covered are (i) Higgs searches (ii)
supersymmetry searches (iii) extra dimensions and (iv) linear collider.
\end{quote}
\bigskip

The projects undertaken in the Working Group I on High Energy and
Collider Physics can be classified into the categories (i) Higgs searches (ii)
supersymmetry searches (iii) extra dimensions and (iv) linear collider. 
The reports on the projects are given below under these headings.

\section{Higgs searches}

\subsection{\boldmath Potential of Associated Higgs Production in LHC through 
$\tau$-pair mode}
\noindent Participants: P.Agrawal and K.Mazumdar
\bigskip

At LHC, the Standard Model Higgs production in association with W-boson,
$pp\rightarrow ~WH$ is very interesting, though in general the total 
production rate is dominated by gluon-gluon fusion process. There is a
strong indication that the Higgs boson is not very heavy and the 
experimental search for Higgs mass, $m_H \le 150$ GeV/c$^2$ is comparatively 
more difficult. In any case it is desirable to study all possibilities 
for detection of Higgs boson in this mass range to strengthen the significance
of discovery via `golden' modes.
This has motivated us to probe less studied modes of Higgs boson 
decays via WH production. Once the Higgs boson is discovered  at LHC 
these final states will have to be studied for confirmation anyway. 

The LHC experiments, both CMS and ATLAS, have special trigger algorithm 
at the first level (LEVEL 1) based on calorimetric information for 
selecting hadronic decays of $\tau$ in the final state ~\cite{CMSTrigger}. 
The $tau$-decay 
modes of Supersymmetric Higgs bosons have been particularly studied for
this purpose. The narrowness of a jet as in the case of hadronic tau decays 
has been utilised in discrimenating the transverse profile of jets. The 
tracker information is used at a later stage for decision at a higher level 
and hence leptonic decays of $\tau$ cannot be used for trigger. Of course the  
leptonic decays of W-boson (only electron and muon final states) can be 
chosen for trigger in inclusive isolated electron/muon mode. But the 
background 
is likely to be overwhelming in that case. Hence we try the possibility
of triggering the signal with the taus from the Higgs decay. This situation 
can be effectively utilised for the decay mode $H\rightarrow \tau^+\tau^-$ 
in the Higgs boson mass range $m_H \le 140$ GeV/c$^2$ where the branching 
ratio is not too small, though below 10\%. 

According to the 'trigger menu' of CMS experiment there are two possibilities 
for events with at least one $\tau$ in the final state. For 95\% efficiency of 
signal selection (SUSY Higgs decay to tau final state) the kinematic 
thresholds are as follows. \\
1.Inclusive $\tau$-jet with jet transverse energy $\ge$ 86 GeV. \\
2. Double $\tau$-jets with the transverse energy of each jet $\ge$ 59 GeV. \\
It remains to be checked through simulation the efficiency in the signal
channel after these requirements. We need to study the spectrum of transverse
momenta of the tau-jets for this.

Assuming that a reasonable fraction of events survive the trigger condition,
we need to reconstruct the events. Since the tau decays will inherently be 
accompanied by missing energy due to the neutrinos, we choose to select the
hadronic decays of W-boson. The W mass can be reconstructed from the jets not 
identified as tau-tagged. Since the taus are highly boosted, the neutrinos 
are expected to be almost collinear with the direction of missing transverse
energy. The mass of the Higgs boson can be reconstructed from this missing
transverse energy and the visible momenta of the tau jets.

The main SM background to this channel is WZ production with 
$Z\rightarrow \tau^+\tau^-$ and $W\rightarrow$ 2-jets. Discarding 
events for which the tau-pair invariant mass is within the Z-mass
window, a good fraction of the background can be removed. We plan to
make a study after detector simulation to evaluate the signal-to-
background ratio. But the Higgs boson being a scalar as opposed to Z, 
some angular correlations  between the tau-jets can be utilised. This 
may not be as easy as in the case with leptons of course and we plan to
make a simulation study of this.

\def\mpT{p_T \hspace{-1em}/\;\:}
\newcommand{\cp}{\mbox{$\not \hspace{-0.15cm} C\!\!P \hspace{0.1cm}$}}

\subsection {Probing the light Higgs window via charged Higgs decay at LHC in 
CP violating MSSM }
\noindent Participants: K. Assamagan, Dilip Kumar Ghosh, 
Rohini M. Godbole  and D.P. Roy
\bigskip

It is well known that {\it all} the observed CP violation in High Energy
Physics can be accommodated in the CKM picture in terms of a single CP-violating
phase. Unfortunately this amount of CP violation in the quark sector, is not
sufficient to explain {\it quantitatively} the observed Baryon Asymmetry in
the Universe. CP violation in the Higgs sector is a popular extension
of the Standard Model, which can cure this deficiency. Of course, CP violation
in the Higgs sector is possible only in Multi-Higgs doublet models, such as
a general two Higgs doublet model (2HDM) or the MSSM. MSSM with complex phases
in the $\mu$ term and soft trilinear SUSY breaking parameters $A_t$
(and $A_b$), can have CP violation in the Higgs sector even with a 
CP-conserving tree level scalar potential. In the presence of these phases,  
due to the CP-violating interactions of the Higgs boson with top and bottom 
squarks, the one loop corrected scalar potential will in general have nonzero 
off-diagonal entries mixing the CP--even (S) and CP--odd (P) states, 
${\cal M}^2_{SP}$, in the $3 \times 3$ neutral Higgs mass-squared matrix. 
After diagonalizing this one-loop corrected scalar potential one will then, 
in general, have three neutral Higgs boson eigenstates, denoted by $H_1, H_2$ 
and $H_3$ in ascending order of masses, with mixed CP parities 
\cite{cpv1,cpv2,cpv3,cpv4,cpv5,cpv6}. 
Sizeable scalar-pseudoscalar mixing is possible 
for large $\mid \mu \mid $ and $\mid A_t \mid ( > M_{SUSY})$.  Such 
CP-violating phases can cause  the Higgs couplings to fermions and gauge
bosons to change significantly from their values at the 
tree-level~\cite{cpv2,cpv4,cpv5}. 

Recently the OPAL Collaboration \cite{opal} has reported their results for the 
Higgs boson searches in the CP-violating MSSM Higgs sector using the 
parameters defined in the CPX scenario \cite{cpv5} using the 
CP-SuperH~\cite{cpsuperh} as well as the FeynHiggs 2.0~\cite{Heinemeyer:2001qd}.
They have provided exclusion regions in the
$M_{H_1}- \tan \beta $ plane for different values of the CP-violating 
phases, assuming ${\rm arg}A_t = {\rm arg}A_b = {\rm arg}M_{\tilde g}  
= \Phi_{\rm CP}$, with $\Phi_{\rm CP} = 90^o, 60^o, 30^o $ and $0^o$. 
The values of the various parameters in the CPX scenario are chosen so 
as to showcase the effects of CP violation in the Higgs sector of the MSSM.
Combining the results of Higgs searches from ALEPH, DELPHI, L3 and OPAL,
the authors in Ref.\cite{cpv7} have also provided exclusion regions in the
$M_{H_1}- \tan \beta $ plane as well as $M_{H^+}- \tan \beta $ plane for 
the above set of parameters. 

Both these analyses show that for phases $\Phi_{\rm CP} = 90^o $ and $60^o$, 
LEP cannot exclude presence of a light Higgs boson for 
$\tan \beta \sim 4-5, M_{H^\pm}\sim 125-140 $ GeV, $ M_{H_1}
\stackrel{<}{{}_\sim} 60 $ GeV and $\tan \beta \sim 2-3, M_{H^\pm}\sim 105-130
 GeV, M_{H_1} \stackrel{<}{{}_\sim} 40 $ GeV respectively. This happens mainly 
due to the reduced  $H_1 ZZ$ coupling, as the lightest Higgs $H_1$ is mostly
a pseudoscalar.  In the same region the $H_1 t \bar t$ 
coupling is suppressed as well. As a result this particular region in the
parameter space can not be probed  at the Tevatron where  the
associated production, $W/Z H_1$ mode is the most promising one; nor can 
it be probed at the LHC as the reduced $t \bar t H_1$ coupling suppresses
the inclusive production mode and  the associated production modes
$W/Z H_1$ and $t \bar t H_1$, are suppressed as well.

It is interesting to note that in the same parameter space where $H_1 ZZ$ 
coupling is suppressed, $H^+ W^- H_1$ coupling is enhanced because these 
two sets of couplings satisfy a sum-rule~\cite{cpsuperh}.
We have found that in these regions of parameter space, $H^\pm \to H_1 W^\pm $
has a very large $(\sim 100 \%) $  branching ratio. This feature motivated us 
to study the possibility of probing such a light Higgs scenario in CP-violating 
MSSM Higgs model 
through the process $p p \to t \bar t \to (b W^\pm) (b H^\mp) \to (b \ell \nu) 
(b H_1 W) \to (b \ell \nu) (b b \bar b ) (jj) $ at LHC. Thus signal will 
consist of 3 or more b-tagged and 2 untagged jets along with a hard lepton 
and missing $p_T$. Similar studies have been done in the context of charged
Higgs search in NMSSSM model \cite{dpr}. 

We report results obtained from a parton level Monte Carlo.
We merge two partons into a single jet if the separation 
$\Delta R = \sqrt{ \left (\Delta \phi \right)^2 +
\left (\Delta \eta \right )^2} < 0.4$.
As a basic selection criteria we require: 
\begin{enumerate}
\item $\mid \eta\mid <2.5 $ for all jets and leptons, where $\eta$ denotes 
pseudo-rapidity, 
\item $p_T$ of the hardest three jets to be higher than 30 GeV,
\item $p_T$ of all the other  jets, lepton, as well as  the missing
$p_T$  to  be larger than 20 GeV, 
\item A minimum separation of $\Delta R = 0.4$ between the lepton and jets 
as well as each pair of jets, 
\item Three or more tagged $b$-jets in the final state  assuming a 
$b$-tagging efficiency of $50\%$. 
\end{enumerate}
\begin{figure}[htbp]
\centerline{\psfig{figure=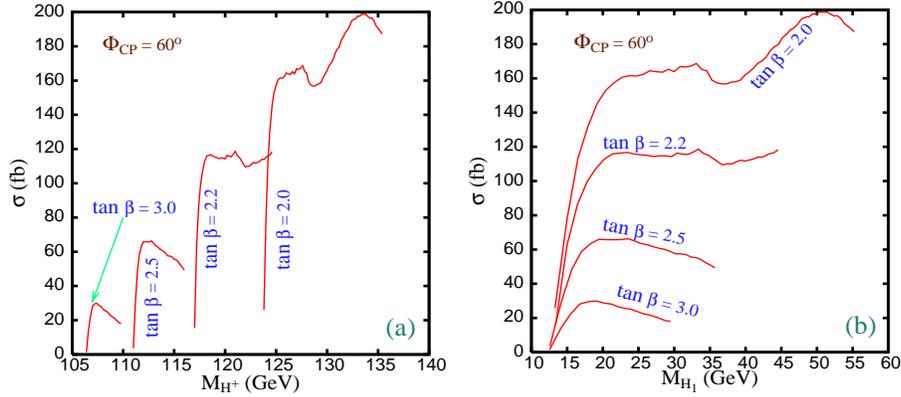,width=12.5cm,height=6cm,angle=0}}
\caption{Variation of signal cross-section with $M_{H^+}$ $(a)$ and $M_{H_1}$
$(b)$ for the CP-violating phase $\Phi_{CP} = 60^o$.}
\label{fig:fig1}
\end{figure}
In Figure~\ref{fig:fig1} we show the 
variation of signal cross-section with $M_{H^+}$ and $M_{H_1}$ for the 
CP-violating phase $\Phi_{\rm CP} = 60^o$. We have used the 
CP-SuperH program~\cite{cpsuperh} to calculate the masses and the couplings
of the Higgses in the CPX scenario.  The cross-section shown in the 
figure includes neither the $b$--tagging efficiency for the three and
more jets (5/16),
nor  the $K$--factor corresponding to the NLO QCD corrections 
for the $t \bar t$ production $(\sim 1.4$--$1.5)$. Hence the numbers in the
figure need to be scaled down by roughly a factor of two  to get the signal 
cross-section.  From Figure~\ref{fig:fig1} one can see
that the signal cross-section decreases with increase in $\tan \beta$. This
can be explained by the fact that $H^+ \to H_1 W^+$ as well as 
$t \to b H^+$ branching ratio decreases with the increase in $\tan \beta$. 
The $t \to b H^+$ branching ratio does increase after showing a dip around 
$\tan \beta \sim 5-6$. However, we are not interested in such a high value of 
$\tan \beta$ in the present investigation as the loss of light Higgs signal 
due to $\cp$ in the Higgs sector is not significant for these higher values of 
$\tan \beta$.

Note that the signal events will be very striking due to the 
clustering of the $b\bar b$, $b\bar b W$ invariant masses at values
corresponding to $M_{H_1}$ and $M_{H^+}$ respectively.  Also the signal
events will have simultaneous clustering of $b \bar b b W$ invariant mass 
around $m_t$.  
\begin{figure}[htbp]
\centerline{\psfig{figure=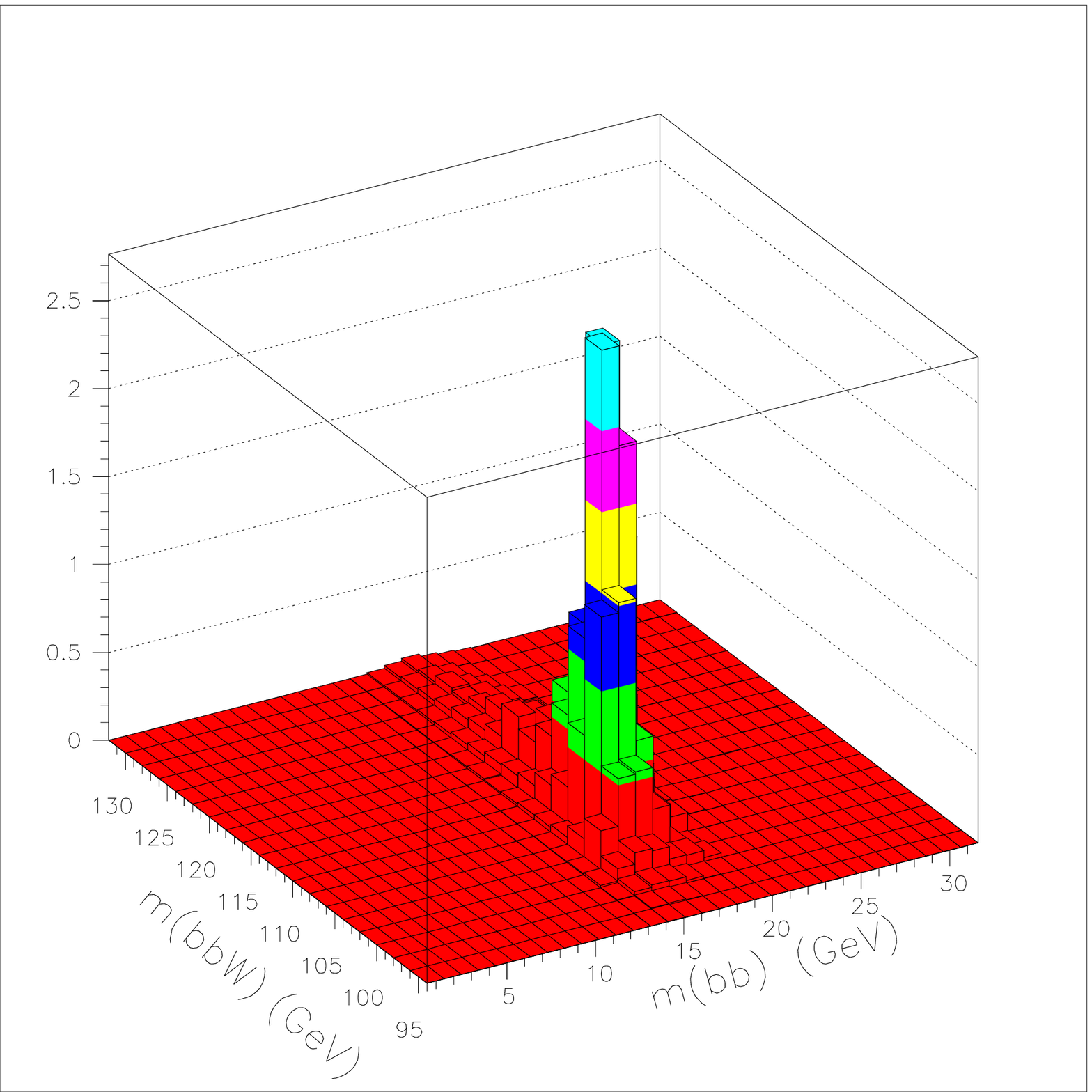,width=5.5cm,height=5.5cm,angle=0}\hspace{1cm}
            \psfig{figure=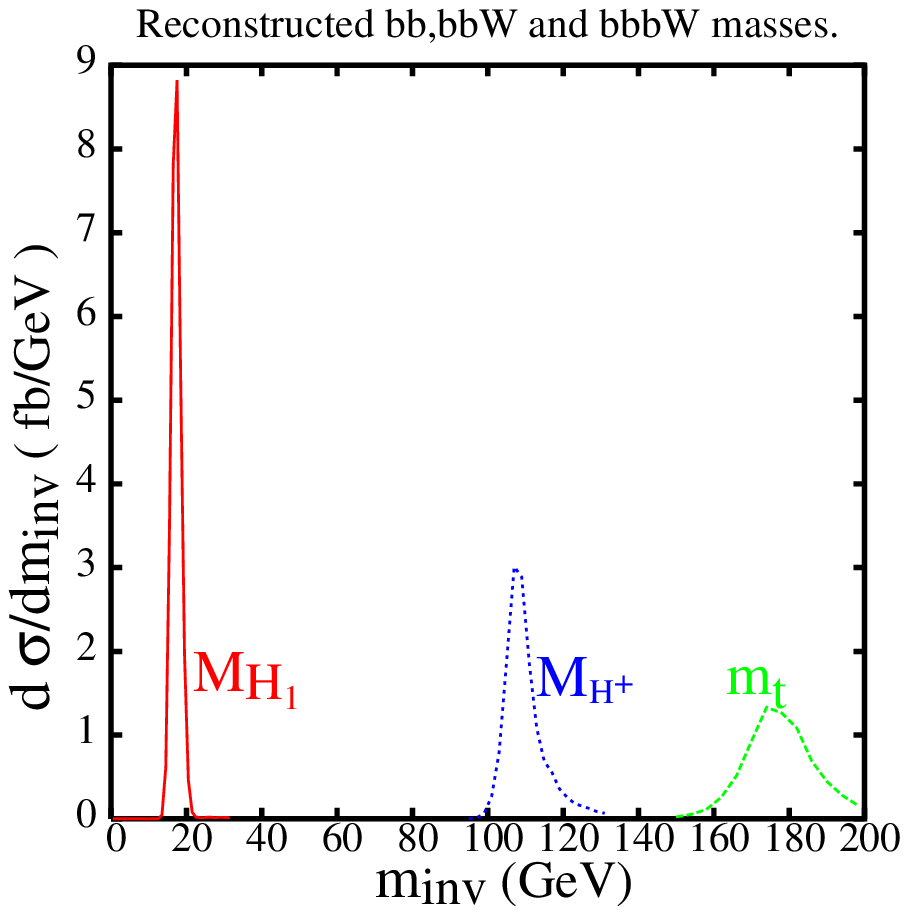,width=6cm,height=6cm,angle=0}
	    } 
\vspace{1cm}
\caption{Clustering of the  $b\bar b$, $b \bar b W $ and $\bar b b b W$
invariant masses around $M_{H_1}$, $M_{H^+}$ and $m_t$. The parameters chosen
for the signal are : CP-violating phase $\Phi_{CP} = 60^o$, $\tan \beta =3$ and
$M_{H^+} = 107$ GeV.}
\label{fig:fig2}
\end{figure}
In Figure~\ref{fig:fig2} we show in the left panel the 
3-dimensional plot for the correlation between $m_{b\bar b}$ and 
$m_{b\bar bW}$ invariant mass distribution for $\Phi_{CP} = 60^o , 
\tan \beta =3 $ and $M_{H^+} = 107$ GeV.  The light Higgs mass corresponding 
to this set of input parameter is 16.78 GeV. It is clear from  
Figure~\ref{fig:fig2} that there is clustering at $M_{H_1}\equiv m_{b\bar b}$ 
and $M_{H^+} \equiv m_{b \bar b W}$.  The right panel of the figure shows the 
same, in terms of cross-section distribution in  $b \bar b$, $\bar b b W$ 
and  $\bar b b b W$  invariant masses for the signal. This makes it very
clear that the detectability of the signal is clearly controlled only by the 
signal size. It is clear from  Figure~\ref{fig:fig1} that indeed the signal
size is healthy over the regions of interest in the parameter space.
The clustering feature can be used to distinguish the signal over the standard 
model background. Thus using this process one can cover, at the LHC, a region 
of the parameter space in $\cp$ MSSM in the $\tan\beta -M_{H_1}$, plane which 
can not be excluded by LEP-2, where the Tevatron has no reach and which
the LHC also can not probe if one does not use the process under 
discussion~\cite{mschumi}.
Of course, in view of the jetty final state, a more rigorous experimental
simulation, including detector effects and hadronisation, will be useful to add further to the strength of our observation. Such a simulation is in 
progress and the results will be presented elsewhere.

\section{Supersymmetry searches}

\def\a               {\alpha}
\def\b               {\beta}
\def\d               {\delta}
\def\g               {\gamma}
\def\G               {\Gamma}
\def\l               {\lambda}
\def\t               {\theta}
\def\s               {\sigma}
\def\x               {\chi}

\def\L               {{\cal L}}
\def\M               {{\cal M}}
\def\H               {{\cal H}}
\def\P               {{\cal P}}

\def\ti              {\tilde}

\def\nt              {\ti\x^0}
\def\ch              {\ti\x^\pm}
\def\sf              {\ti f}
\def\stau            {\ti \tau}
\def\stop            {\ti t}
\def\sbottom         {\ti b}
\def\staun           {\ti \nu_\tau}
\def\tsf             {\theta_{\ti f}}

\def\phmu            {\phi_\mu}
\def\phA             {\phi_A}
\def\phsf            {\varphi_{\!\sf}^{}}
\def\phst            {\varphi_{\ti t}^{}}
\def\phsb            {\varphi_{\ti b}^{}}
\def\phstau          {\varphi_{\ti\tau}^{}}

\newcommand{\lsp}    {m_{\ti \x^0_1}}
\newcommand{\mnt}[1] {m_{\ti \x^0_{#1}}}
\newcommand{\mch}[1] {m_{\ti \x^\pm_{#1}}}
\newcommand{\msf}[1] {m_{\sf_{#1}}}
\newcommand{\mst}[1] {m_{\ti t_{#1}}}
\newcommand{\msb}[1] {m_{\ti b_{#1}}}
\newcommand{\msl}[1] {m_{\ti\tau_{#1}}}

\def\PL              {P_L^{}}
\def\PR              {P_R^{}}

\def\fb              {${\rm fb}^{-1}$}
\def\gev             {{\rm GeV}}
\def\rzw             {\sqrt{2}}
\def\delr            {\!\stackrel{\leftrightarrow}{\partial^\mu}\!}
\def\to              {\rightarrow}

\newcommand{\nn}{\nonumber}
\newcommand{\noi}{\noindent}

\newcommand{\mbf}      {\boldmath}
\newcommand{\sfrac}[2] {{\textstyle \frac{#1}{#2}}}
\newcommand{\smaf}[2]  {{\textstyle \frac{#1}{#2} }}
\newcommand{\pif}      {\smaf{\pi}{2}}

\newcommand{\eq}[1]  {\mbox{(\ref{eq:#1})}}
\newcommand{\fig}[1] {\mbox{Fig.~\ref{fig:#1}}}
\newcommand{\figs}[1] {\mbox{Figs.~\ref{fig:#1}}}
\newcommand{\Fig}[1] {\mbox{Figure~\ref{fig:#1}}}
\newcommand{\sect}[1] {\mbox{Section~\ref{sect:#1}}}
\newcommand{\Sect}[1] {\mbox{Section~\ref{sect:#1}}}

\renewcommand{\Re}{{\cal R}e\,}
\renewcommand{\Im}{{\cal I}m\,}


\newcommand{\gsim}{\;\raisebox{-0.9ex}
           {$\textstyle\stackrel{\textstyle >}{\sim}$}\;}
\newcommand{\lsim}{\;\raisebox{-0.9ex}{$\textstyle\stackrel{\textstyle<}
           {\sim}$}\;}


\subsection{Fermion polarization in sfermion decays
as a probe of CP phases in the MSSM}
\noindent Participants: Thomas~Gajdosik, Rohini M. Godbole and Sabine~Kraml
\bigskip

\noindent {\bf Introduction}
CP violation is one feature of the SM that still defies a fundamental
theoretical understanding, even though {\it all} 
the observed CP violation in High Energy Physics can be accommodated in the 
CKM picture in terms of a single CP-violating phase. However, this amount 
of CP violation in the quark sector is not sufficient to explain 
{\it quantitatively} the observed Baryon Asymmetry in the Universe. 
CP violation in the Higgs sector is a popular extension
of the Standard Model, which might cure this deficiency. 
MSSM with complex phases in the $\mu$ term and soft trilinear SUSY 
breaking parameters $A_t$ (and $A_b$), can affect the Higgs sector
~\cite{Dedes:1999sj,Carena:2002bb}  through loop corrections.
One can then have CP-violating effects 
even with a CP-conserving tree level scalar potential. It is still possible
to be  
consistent with the non-observation of the electron EDMs (eEDM). 
This makes the MSSM with CP-violating phases a very attractive
proposition. It has therefore been the subject of many recent
investigations, studying the implications of these phases on
neutralino/chargino production and decay~\cite{Bartl:2004ut},
on the third generation of sfermions~\cite{Bartl:2003he} 
as well as the neutral~\cite{Carena:2002bb,Borzumati:2004rd}
and charged \cite{Christova:2002ke} Higgs sector.  
In these studies, the gaugino mass parmaeter $M_1$ is also taken to be 
complex in addition to the nonzero phases mentioned above.
It is interesting to 
note that CP-even observables such as masses, branching ratios, cross
sections, etc., often afford more precise probes of these phases, 
thanks to the larger magnitudes of the effects as compared to the 
CP-odd/T-odd observables. The latter, however, are the only ones that 
can offer direct evidence of CP violation~\cite{Bartl:2004ut}.  A recent 
summary of the progress in the area can be found in~\cite{heselbach,ggk1}
and references therein.

In this project, we address the issue of probes of these phases through a study
of the third generation sfermions.  A recent study in this context, in 
the $\stop, \sbottom$ sector in  the second of Ref.~\cite{Bartl:2003he},
demonstrates that it may be possible to determine the real and imaginary
parts of $A_t (A_{\tau})$ to a precision of 2--3\% (10--20 \% for low 
$\tan \beta$ and 3--7\% at large $\tan \beta$)  from a fit of the MSSM Lagrange
parameters to masses, cross sections and branching ratios at a future LC.
In this project~\cite{ggk1} we have explored the the longitudinal polarization 
of fermions produced in sfermion decays, i.e.\ $\ti f\to f\nt$ and
$\ti f\to f'\ch$ with $f(\ti f)$ a third generation (s)quark or
(s)lepton, as a probe of CP phases.

The average polarization of fermions produced in sfermion decays carries 
information on the $\ti f_L^{}$--$\ti f_R^{}$ mixing as well as on the 
gaugino--higgsino mixing \cite{Nojiri:1994it}. The polarizations that can 
be measured are those of top and tau; both can be inferred from the decay 
product(lepton angle  and/or pion energy) distributions.  The use of 
polarization of the decay fermions for studies of MSSM 
parameter determination was first pointed out and demonstrated in 
Ref.~\cite{Nojiri:1994it,Nojiri:1996fp}. An extension of these ideas 
for the CP-violating case and the phase dependence of the longitudinal 
fermion polarization had been mentioned in the studies of \cite{heselbach}.
We provide, in this note, a detailed discussion of the sensitivity
of the fermion polarization to the CP-violating phases in the MSSM.

\smallskip
\noindent {\bf  Fermion polarization in $\sf\to f\nt$ decays}

\noi
The sfermion interaction with neutralinos is ($i=1,2$; $n=1,...,4$)
\begin{eqnarray}
  \L_{f\sf\nt}
  &=& g\,\bar f\,( a^{\,\sf}_{in}\PR + b^{\,\sf}_{in}\PL )\,\nt_n\,\sf_i^{}
      + {\rm h.c.}
\end{eqnarray}
Thus $ a^{\,\sf}_{in} (b^{\,\sf}_{in}) $ determine the amplitude for the
production of $f_L (f_R)$ in the decay $\sf_{i}\to f\nt_{n}$.  
\noi
The gaugino interaction conserves the helicity of the
sfermion while the higgsino interaction flips it. In the limit
$m_f\ll\msf{i}$, the average polarization of the fermion coming
from the above  decay can therefore be calculated as \cite{Nojiri:1994it}
\begin{equation}
   \P_f^{} = \frac{Br(\sf_i\to\nt_n f_R)-Br(\sf_i\to\nt_n f_L)}
                  {Br(\sf_i\to\nt_n f_R)+Br(\sf_i\to\nt_n f_L)}
           = \frac{|b_{in}^{\sf}|^2-|a_{in}^{\,\sf}|^2}
                  {|b_{in}^{\sf}|^2+|a_{in}^{\,\sf}|^2} \,.
   \label{eq:Pf}
\end{equation}
\noi
We obtain  for the $\sf_1^{}\to f\nt_n$
decay (omitting the overall factor $g^2$ and dropping the sfermion
and neutralino indices for simplicity):
\begin{eqnarray}
   |b_{1n}^{}|^2-|a_{1n}^{}|^2
   &=& |h_L^{}\cos\t\,e^{-i\varphi} + f_R^{}\sin\t|^2 -
       |f_L^{}\cos\t\,e^{-i\varphi} + h_L^*\sin\t|^2    \nn\\
   &=& (|h_L|^2-|f_L|^2)\cos^2\t - (|h_L|^2-|f_R|^2)\sin^2\t \nn\\
   & & + \,\sin 2\t\,\big[\,
       \Re (f_R^{}-f_L^{})\,(\Re h_L^{}\cos\varphi+\Im h_L^{}\sin\varphi) \nn\\
   & & \hspace*{14mm}
       +\,\Im (f_R^{}+f_L^{})\,(\Im h_L^{}\cos\varphi-\Re h_L^{}\sin\varphi)
       \,\big] ,
\end{eqnarray}
where $\t, \varphi$ are the sfermion mixing angle and phase, and
$f_L,f_R$ and $h_L,h_R$ are the gaugino and higgsino couplings of the left-
and right-chiral sfermions respectively\footnote{For details see \cite{ggk1}.}
and contain the dependence on the phases in the gaugino--higgsino sector, 
$\phi_1, \phi_\mu$.
We see that the phase dependence of $\P_f^{}$ is the largest for
maximal sfermion mixing ($\tsf=3\pi/4$) and if the neutralino has
both sizeable gaugino and higgsino components. It is, moreover,
enhanced if the Yukawa coupling $h_f$ is large.
Furthermore, $\P_f^{}$ is sensitive to CP violation even if just
one phase, in either the neutralino or the sfermion sector, is
non-zero. In particular, if only $A_f$ and thus only the sfermion
mixing matrix has a non-zero phase, the phase-dependent term
becomes
\begin{equation}
   |b_{1n}^{}|^2-|a_{1n}^{}|^2
   \: \stackrel{\phi_1=\phi_\mu=0}{\sim} \:
   h_L^{}(f_L^{}-f_R^{})\sin2\t\cos\varphi \,.
\end{equation}
    The polarization $\P_{\!f}^{}$, eq.~\eq{Pf}, depends only on
couplings but not on masses. For the numerical analysis we
therefore use $M_1$, $M_2$, $\mu$, $\tan\beta$, $\tsf$ and $\phsf$
as input parameters, assuming $\phmu \approx 0$ to satisfy EDM
constraints more easily: assuming cancellations for the 1-loop
contributions and the CP-odd Higgs mass parameter $m_{A} > 300$~GeV, 
1-loop and 2-loop contributions to the electron EDM (eEDM), 
as well as their sum, stay below the experimental limit~\cite{Chang:1998uc,ggk1}
    In order not to vary too many parameters, we use,
moreover, the GUT relation $|M_1|=\frac{5}{3}\tan^2\theta_W M_2$
and choose $\tan\b=10$ and $\t_{\ti t}=\t_{\ti\tau}=130^\circ$;
i.e., 
large but not maximal mixing. The free parameters in this analysis are 
thus $M_2$, $|\mu|$, and the phases $\phi_1$, $\phsf$.
\begin{figure}[htb]
\centerline{\psfig{figure=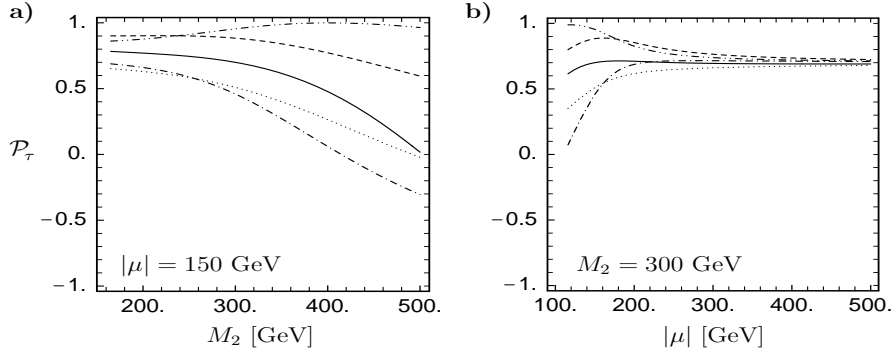,width=13cm,height=5cm,angle=0}}
\caption{Average polarization of the tau lepton coming from
$\ti\tau_1\to\tau\nt_1$ decays in a) as a function of $M_2$ 
, in b) as a function of $|\mu|$. 
The full, dashed, dotted, dash-dotted, and dash-dot-dotted
lines are for $(\phi_1,\,\varphi_{\ti \tau})=(0,\,0)$,
$(0,\,\frac{\pi}{2})$, $(\frac{\pi}{2},0)$,
$(\frac{\pi}{2},\,\frac{\pi}{2})$, and
$(\frac{\pi}{2},\,-\frac{\pi}{2})$, respectively.
$M_2$ and $\mu$ are taken to be real.
\label{fig:Ptau_M2mu}}
\end{figure}
\Fig{Ptau_M2mu} shows the average tau polarization in
$\ti\tau_1\to\tau\nt_1$ decays as functions of $M_2$ and $|\mu|$, for values 
consistent with the LEP constraints,
for $\tan\b=10$, $\theta_{\ti\tau}=130^\circ$ and various choices
of $\phi_1$ and $\varphi_{\ti\tau}$.  We find that the 
$\P_{\!\tau}^{}$ is quite sensitive to CP phases for $|\mu|<M_2$,
when $\nt_1$ has a sizeable higgsino component.  Similarly 
the average top polarization
in $\ti t_1\to t\nt_1$ decays can be studied.  We find that not only
does it  have a strong dependence on the 
CP phases if the neutralino has a sizeable higgsino component, but it is 
also
significant when $|\mu|\sim M_2$, due to the much larger value of $m_t$ 
compared to $m_\tau$. Since at a future $e^+e^-$ linear collider (LC), 
one expects to be able to measure the tau polarization to about
3--5\% and the top polarization to about 10\% \cite{Boos:2003vf},
the effects of CP-violating phases may well be visible in
$\P_{\!t}^{}$ and/or $\P_{\!\tau}^{}$, provided $\mu$ is not too
large.

\begin{figure}[htb]
\centerline{\psfig{figure=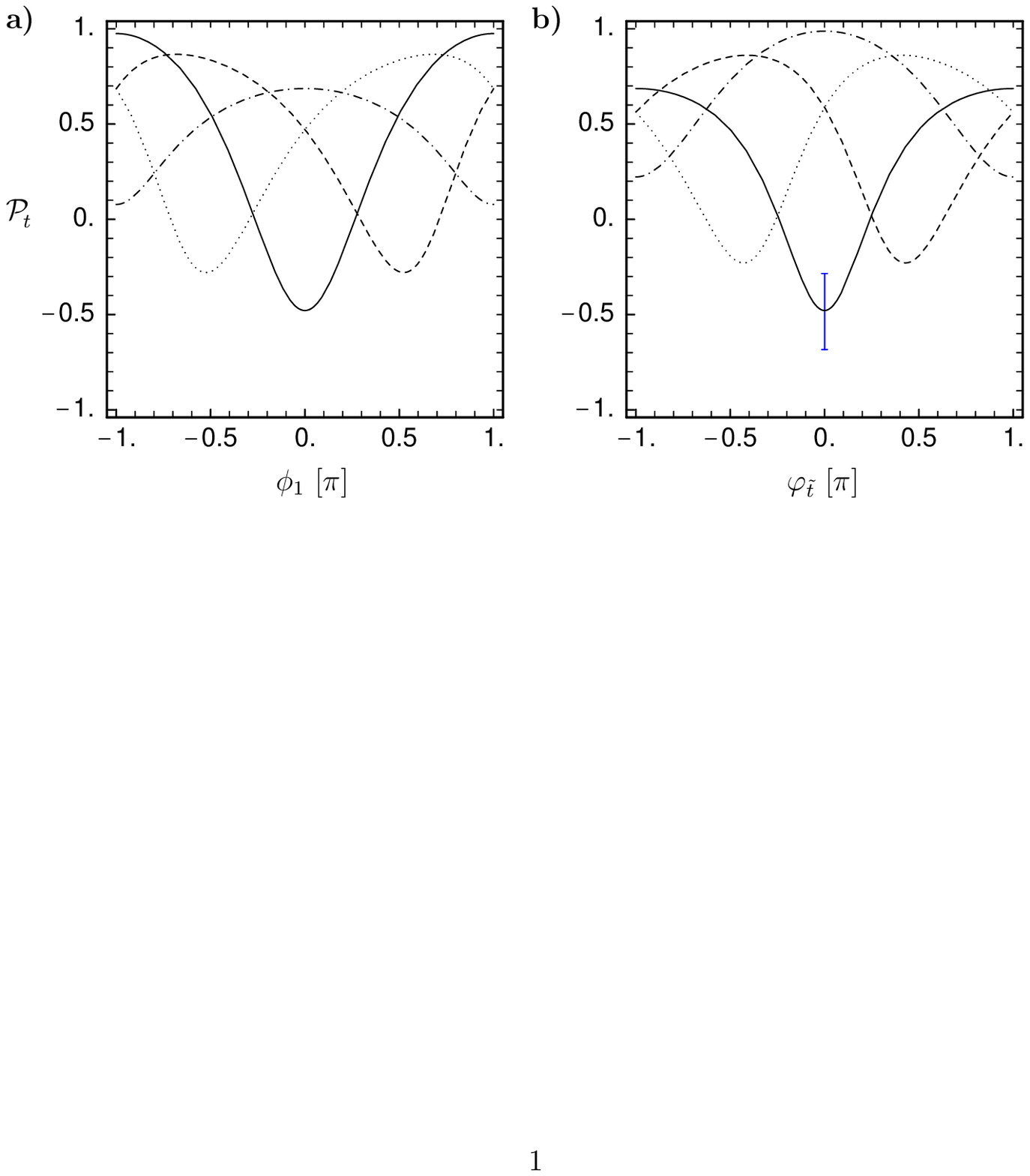,width=13cm,height=5cm,angle=0}}
\caption{Average polarization of the top quark coming from $\ti
t_1\to t\nt_1$ decays for $\theta_{\ti t}=130^\circ$, and
$\tan\beta=10$: in a) as a function of $\phi_1$ for $M_2=225$~GeV
and $|\mu|=200$~GeV; in b) as a function of $\varphi_{\ti t}$ for
$|\mu|=200$~GeV and $M_2$ adjusted such that $\mnt{1}=100$~GeV.
The full, dashed, dotted, and dash-dotted lines are for $\phst$
($\phi_1$) $= 0,\pif,-\pif,\pi$ in a (b). The error on
$\P_{\!t}^{}$ indicated by the vertical bar in b) has been
estimated as described in the text. \label{fig:Ptop_phases}}
\end{figure}

The phase dependence is further studied in \Fig{Ptop_phases} where we show 
$\P_{\!t}^{}$ as a function of $\phi_1$, for $M_2=380$~GeV,
$|\mu|=125$~GeV and $\phst=0$, $\frac{\pi}{2}$, $-\frac{\pi}{2}$
and $\pi$. Since for fixed $M_2$ and $|\mu|$ the $\nt_1$ mass
changes with $\phi_1$, we show in addition in \fig{Ptop_phases}b
$\P_{\!t}^{}$ as a function of $\phst $ for various values of
$\phi_1$, with $|\mu|=125$~GeV and $M_2$ adjusted such that
$\mnt{1}=100$~GeV.
We thus see that if the neutralino mass parameters, $\tan\b$ and
$\t_{\ti t}$ are known, $\P_{\!t}^{}$ can hence be used as a
sensitive probe of these phases (although additional information
will be necessary to resolve ambiguities and actually determine
the various phases).
The influence of uncertainties in the knowledge of
the SUSY model parameters,  can be studied  by choosing 
the case of $M_2=380$~GeV,
$|\mu|=125$~GeV and vanishing phases as reference point and assume
that the following precisions can be achieved:
$\d M_1=\d M_2=\d\mu=0.5\%$,
$\d\tan\b=1$, $\d\t_{\ti t}=3.5^\circ$, and
$\d\phi_1=\d\phi_\mu=0.1$. Varying the parameters within this
range around the reference point and adding experimental
resolution $\d\P_{\!t}^{exp}\simeq 0.1$ in quadrature gives
$\P_{\!t}^{}=-0.48\pm 0.22$ at $\phst=0$, indicated as an error
bar in \fig{Ptop_phases}b. The figure shows that in  this scenario
$\P_{\!t}^{}$ would be sensitive to $|\phst|\gsim 0.15\pi$. 
If more accurate measurements of the SUSY parameters should be available
such that 
$\d\P_{\!t}^{par}$ would be negligible
compared to the experimental resolution of $\P_{\!t}$, 
then it would be possible to derive information 
on $A_t$ 
using the $\P_{\!t}^{}$ measurement.

\smallskip
\noindent
{\bf Fermion polarization in $\sf\to f'\ch$ decays} 

\noi
Analogous to the decay into a neutralino, eq.~\eq{Pf}, the average
polarization of the fermion coming from the $\sf_i\to f'\ch_j$ decay
($i,j=1,2$) can be calculated once we know the $\sf_i f' \ch_j$ 
coupling.  These can be read off from the interaction Lagrangian:  
\begin{eqnarray}
  \L_{f'\!\sf\ch}
  &=& g\,\bar u\,( l_{ij}^{\,\ti d}\,\PR +
                   k_{ij}^{\,\ti d}\,\PL )\,\ti\x^+_j\,\ti d_i^{}
    + g\,\bar d\,( l_{ij}^{\,\ti u}\,\PR +
                   k_{ij}^{\,\ti u}\,\PL )\,\ti\x^{+c}_j\,\ti u_i^{}
      + {\rm h.c.}
\end{eqnarray}
where $u$ ($\ti u$) stands for up-type (s)quark and (s)neutrinos,
and $d$ ($\ti d$) stands for down-type (s)quark and charged (s)leptons.
The average polarization is then given by
\begin{equation}
   {\P_{\!f}^{}}'
           = \frac{Br\,(\sf_i\to\ch_j f_R')-Br\,(\sf_i\to\ch_j f_L')}
                  {Br\,(\sf_i\to\ch_j f_R')+Br\,(\sf_i\to\ch_j f_L')}
           = \frac{|k_{ij}^{\sf}|^2-|l_{ij}^{\,\sf}|^2}
                  {|k_{ij}^{\sf}|^2+|l_{ij}^{\,\sf}|^2} \,.
\label{eq:Pfprime}
\end{equation}
\noindent
Since only top and tau polarizations are measurable, we
studied $\ti b\to t\ti\x^-$ and $\ti\nu_\tau\to\tau\ti\x^+$
decays. The latter case is especially simple because
${\P_{\!\tau}^{}}'$ depends only on the parameters of the chargino
sector:
A measurement of ${\P_{\!\tau}^{}}'$ may hence be useful to
supplement the chargino parameter determination.  However, only for the
decay into the heavier chargino, the effect of a non-zero
phase
may be sizeable.  
Recall that unless huge cancellations
are invoked, $\phi_\mu^{}$ is severely restricted by the
non-observation of the eEDM. 
Moreover, the measurement
of $({\P_{\!\tau}^{}}')_2^{}$ will be diluted by
$\ti\nu_\tau\to\tau\ti\x^+_1$ .

The  top polarization in $\ti b\to t\ti\x^-$ decays is more promising.
Again we find  that the phase dependence of ${\P_{\!t}^{}}'$ is
proportional to $h_b \sin 2\theta_{\ti b}$ and the amount of
gaugino--higgsino mixing of the charginos; it will therefore be
largest for $|M_2|\sim|\mu|$, $\theta_{\ti b}=3\pi/4$ and large
$\tan\beta$. Again, there is a non-zero effect even if there is
just one phase in either the sbottom or chargino sector. Note,
however, that the only CP phase in the chargino sector
is $\phi_\mu$, which also enters the sfermion mass matrices.
As a result, depending on values of $A_b, \tan \beta$ and $\mu$,
$\phsb$ and $\phi_\mu^{}$ get related.
For the sake of a general discussion of the phase dependence
of ${\P_{\!t}^{}}'$ (and since $A_b$ is still a free parameter),
we nevertheless use $\phi_\mu^{}$ and $\phsb$ as independent
input parameters.  If
$\phi_\mu^{}$ and $\phsb$ have the same sign, the difference in
${\P_{\!t}^{}}'$ from the case of vanishing phases is larger than
if they have opposite signs. In particular, we find
${\P_{\!t}^{}}'(\phi_\mu=-\varphi_{\ti b})\sim
{\P_{\!t}^{}}'(\phi_\mu=\varphi_{\ti b}=0)$ over large regions of
the parameter space. With an experimental resolution of the top
polarization of about 10\% this implies that in many cases
$\varphi_{\ti b}\sim -\phi_\mu$ cannot be distinguished from
$\varphi_{\ti b}=\phi_\mu=0$ by measurement of ${\P_{\!t}^{}}'$.

\begin{figure}[htb]
\centerline{\psfig{figure=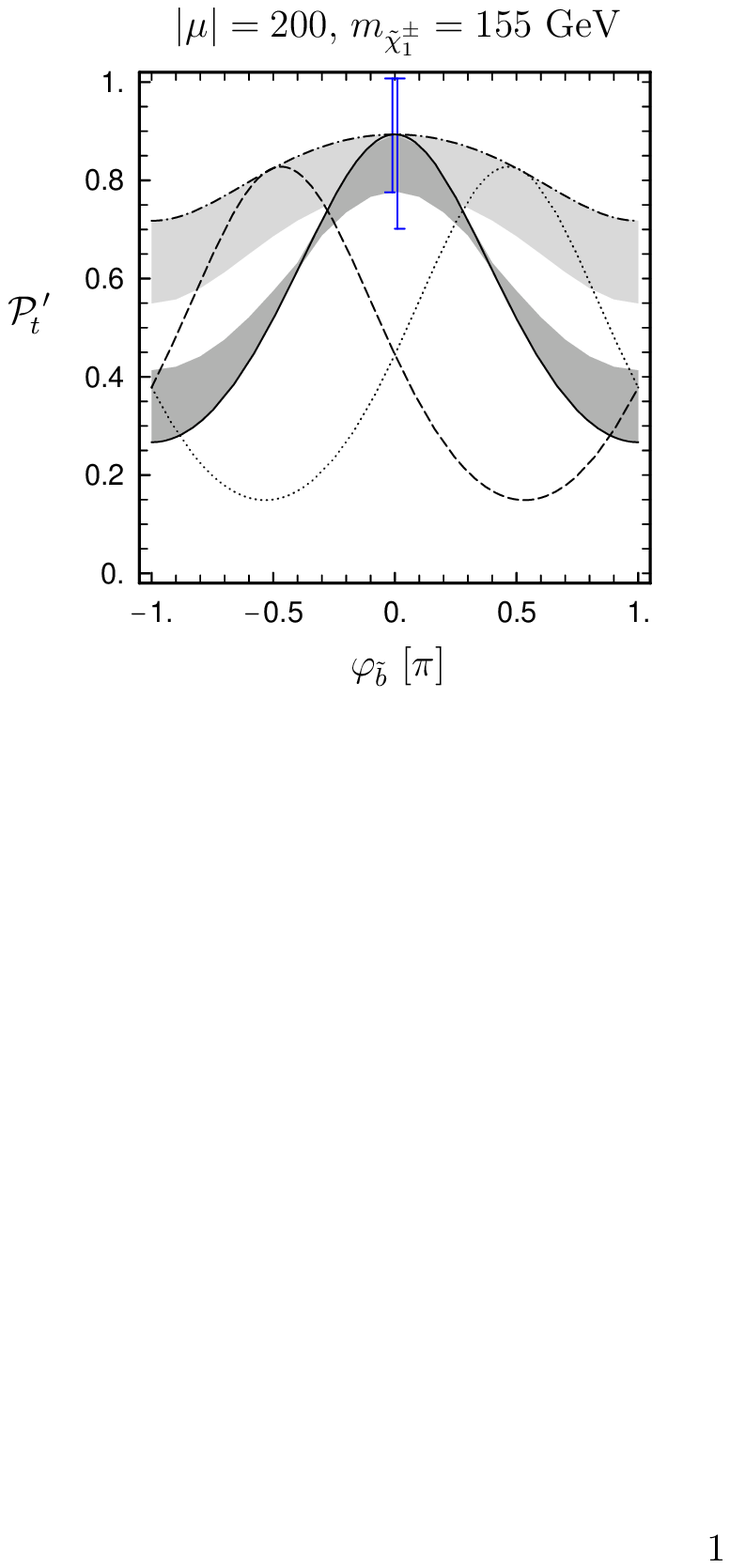,width=7cm,height=5cm,angle=0}}
\caption{Average polarization of the top quark coming from $\ti
b_1^{}\to t\ti\x^-_1$ decays as a function of $\varphi_{\ti b}$,
The full, dashed and dotted lines are for $\phi_\mu = 0$,
$\frac{\pi}{2}$ and $-\frac{\pi}{2}$, respectively, while for the
dash-dotted lines $\phi_\mu = -\varphi_{\ti b}$. The grey bands
show the range of ${\P_{\!t}^{}}'$ due to varying $m_b$ within
2.5--4.5 GeV for the cases $\phi_\mu=0$ and $\phi_\mu =
-\varphi_{\ti b}$. The error bars show the estimated errors on
${\P_{\!t}^{}}'$ as described in the text. \label{fig:Ptoppr_phi}}
\end{figure}

As an example of the phase dependence of the polarization ${\P_{\!t}^{}}'$
we show some of our results  in \fig{Ptoppr_phi}
which shows ${\P_{\!t}^{}}'$ as a function of $\phsb$,
for $|\mu|=200$~GeV, $\tan\beta=10$, $\theta_{\ti b}=140^\circ$,
and various values of $\phi_\mu$. $M_2$ is chosen such that
$\mch{1}=155$~GeV (i.e.\ $M_2=225$~GeV for $\phi_\mu=0$). The
range obtained by varying $m_b$ within 2.5--4.5~GeV is shown as
grey bands for two of the curves, for $\phi_\mu=0$ and
$\phi_\mu=-\phsb$. We estimate the effect of an imperfect
knowledge of the model parameters in the same way as in the
previous section. For $M_2=225\pm 1.125$~GeV, $|\mu|=200\pm
1$~GeV, $\tan\b=10\pm 1$, $\t_{\ti b}=140\pm 3.4^\circ$ and
$\phi_\mu=0\pm 0.1$, we get ${\P_{\!t}^{}}'=0.89 \pm 0.06$ at
$\phsb=0$. Varying in addition $m_b=2.5$--4.5 GeV gives
${\P_{\!t}^{}}'=0.89^{+0.06}_{-0.16}$. Adding a 10\% measurement
error on ${\P_{\!t}^{}}'$ in quadrature, we end up with
${\d\P_{\!t}^{}}'=0.12$ ($0.19$) without (with) the $m_b$ effect.
These are shown as error bars in \fig{Ptoppr_phi}.
We see that the case of $\phsb=-\phi_\mu^{}$ cannot be
distinguished from $\phsb=\phi_\mu^{}=0$ in this scenario.
However, ${\P_{\!t}^{}}'$ turns out to be quite a sensitive probe
of $\delta_\phi^{} = \phsb+\phi_\mu^{}$, i.e.\ the deviation from
the `natural' alignment $\phsb=-\phi_\mu^{}$. In the example of 
 \fig{Ptoppr_phi}, $|\delta_\phi^{}|\gsim 0.24\pi$ $(0.31\pi)$ can
be resolved if $h_b$ is (not) known precisely, quite independently
of $\phi_\mu^{}$. Observing such a $\delta_\phi^{}$ also implies a
bound on $|A_b|$ of $|A_b| > 1363$ $(1678)$~GeV. If the precision
on $M_2$ and $|\mu|$ is 0.1\% and $\tan\b=10\pm 0.1$, we get
$({\d\P_{\!t}^{}}')^{par}= 0.03$ at $\phsb=0$, so that the error
is dominated by the experimental uncertainty. However, the
resultant improvement in the sensitivity is limited to
$|\delta_\phi^{}|\gsim 0.22\pi$ and $|A_b| > 1294$~GeV.

\vspace{0.3cm}

\noi
{\bf Summary}

\noi We have investigated the sensitivity of the longitudinal
 polarization of fermions (top and tau) produced in sfermion decays 
 to CP-violating phases in the MSSM. We have found that both
 $\P_{\!t}^{}$ and $\P_{\!\tau}^{}$ can vary over a large range
 depending on $\phi_1$ and $\varphi_{\ti t,\ti\tau}$ (and also
 $\phi_\mu$, though we did not discuss this case explicitly) and
 may thus be used as sensitive probes of these phases. To this aim,
 however, the neutralino mass parameters, $\tan\b$ and the sfermion
 mixing angles need to be known with high precision. Given the 
complexity 
 of the problem, a combined fit of all available data seems to be the 
most
 convenient method for the extraction of the MSSM parameters. For
 the decays into charginos, the tau polarization in
 $\ti\nu_\tau\to\tau\ti\x^+$ decays depends only little on
 $\phi_\mu^{}$. ${\P_{\!\tau}^{}}'$ is hence not a promising
 quantity to study CP phases, but may be useful for (consistency)
 tests of the gaugino--higgsino mixing. The top polarization in
 $\ti b\to t\ti\x^-$ decays, on the other hand, can be useful to
 probe $\phi_\mu^{}$, $\phsb$ and/or $\d_\phi^{}=\phi_\mu^{}+\phsb$
 in some regions of the parameter space. The measurement of
 ${\P_{\!t}^{}}'$, revealing phases or being consistent with
 vanishing phases, may also constrain $|A_b|$. For a more detailed
 report of our investigations see~\cite{ggk1}.

\newcommand{\newc}{\newcommand}
\def\preprint{{preprint}}
\def\Ord{\lower .7ex\hbox{$\;\stackrel{\textstyle <}{\sim}\;$}}
\def\OOrd{\lower .7ex\hbox{$\;\stackrel{\textstyle >}{\sim}\;$}}
\def\cO#1{{\cal{O}}\left(#1\right)}
\newc{\anti}{\overline}
\newc{\lra}{\longrightarrow}
\newc{\wt}{\widetilde}
\newc{\tim}{\times}
\newc{\lam}{\lambda}
\newc{\Lam}{\Lambda}
\newc{\gam}{\gamma}
\newc{\Gam}{\Gamma}
\newc{\eps}{\epsilon}
\newc{\Eps}{\Epsilon}
\newc{\kap}{\kappa}
\newc{\modulus}[1]{\left| #1 \right|}
\newc{\eqs}[2]{(\ref{eq:#1},\ref{eq:#2})}
\newc{\ibid}{{\it ibid}.}
\newc{\eg}{{\it e.g.}\ }
\newc{\ie}{{\it i.e.}\ }
\def \viz{\emph{viz.}}
\def \etc{\emph{etc. }}
\newc{\nonum}{\nonumber}
\newc{\lab}[1]{\label{eq:#1}}
\newc{\dpr}[2]{({#1}\cdot{#2})}
\newc{\lsimeq}{\stackrel{<}{\sim}}
\newc{\lt}{\stackrel{<}}
\newc{\gt}{\stackrel{>}}
\newc{\gsimeq}{\stackrel{>}{\sim}}
\def\lsim{\buildrel{\scriptscriptstyle <}\over{\scriptscriptstyle\sim}}
\def\gsim{\buildrel{\scriptscriptstyle >}\over{\scriptscriptstyle\sim}}
\newc{\half}{\frac{1}{2}}
\def\bra{\langle}
\def\ket{\rangle}
\def\cO#1{{\cal{O}}\left(#1\right)}
\def \DM{{\Delta{m}}}
\def\lapp{\mathrel{\rlap{\raise.5ex\hbox{$<$}}
                    {\lower.5ex\hbox{$\sim$}}}}
\def\gapp{\mathrel{\rlap{\raise.5ex\hbox{$>$}}
                    {\lower.5ex\hbox{$\sim$}}}}
\newc{\dota}{\dot{\alpha }}
\newc{\dotb}{\dot{\beta }}
\newc{\dotd}{\dot{\delta }}
\newc{\del}{\delta}
\newc{\Del}{\Delta}
\newc{\nindnt}{\noindent}
\newc{\matth}{\mathsurround=0pt}
\def\ML{\ifmmode{{\mathaccent"7E M}_L}
             \else{${\mathaccent"7E M}_L$}\fi}
\def\MR{\ifmmode{{\mathaccent"7E M}_R}
             \else{${\mathaccent"7E M}_R$}\fi}

\def \ud { {1 \over 2} }
\def \ut { {1 \over 3} }
\def \td { {3 \over 2} }
\newc{\mr}{\mathrm}

\def\sig{\sigma}
\def\Sig{\Sigma}
\def\dh {\partial }
\newc{\delgs}{\delta_{GS}}
\def \cc { coupling constant }
\def \ccs {coupling constants }
\def \ps {parameter space }
\def \pps {parameter spaces }


\newc{\thetaw}{\theta_W}
\def \PI{{\pi^{\pm}}}
\newc{\ppbar}{\mbox{$p\overline{p}$}}
\newc{\bbbar}{\mbox{$b\overline{b}$}}
\newc{\ccbar}{\mbox{$c\overline{c}$}}
\newc{\ttbar}{\mbox{$t\overline{t}$}}
\newc{\eebar}{\mbox{$e\overline{e}$}}
\newc{\zzero}{\mbox{$Z^0$}}
\def \gamz{\Gamma_Z}
\newc{\wplus}{\mbox{$W^+$}}
\newc{\wminus}{\mbox{$W^-$}}
\newc{\elp}{\mbox{$e^+$}}
\newc{\elm}{\mbox{$e^-$}}
\newc{\elpm}{\mbox{$e^{\pm}$}}
\newc{\qbar}     {\mbox{$\overline{q}$}}
\newc{\lp}{\mbox{$e^+$}}
\newc{\lm}{\mbox{$e^-$}}
\newc{\lpm}{\mbox{$e^{\pm}$}}
\def \ewgroup{SU(2)_L \otimes U(1)_Y}
\def \smgroup{SU(3)_C \otimes SU(2)_L \otimes U(1)_Y}


\def \poincare{Poincare$\acute{e}$}
\def \superspace{\emph{superspace}}
\def \sf{\emph{superfield}}
\def \sfs{\emph{superfields}}
\def \superpot{\emph{superpotential}}
\def \csf{\emph{chiral superfield}}
\def \csfs{\emph{chiral superfields}}
\def \vsf{\emph{vector superfield }}
\def \vsfs{\emph{vector superfields}}
\newc{\Ebar}{{\bar E}}
\newc{\Dbar}{{\bar D}}
\newc{\Ubar}{{\bar U}}
\def \PROCESS{e^+e^- \rightarrow \wt{\chi}^+ \wt{\chi}^- \gamma}

\newc{\gluino}   {\mbox{$\wt{g}$}}
\newc{\mgl}  {\mbox{$m(\gluino)$}}


\def \charginopm{{\wt\chi}^{\pm}}
\def \mcharginopm{m_{\charginopm}}
\def \mchpmmin {\mcharginopm^{min}}
\def \choneplus {{\wt\chi_1^+}}
\def \choneminus {{\wt\chi_1^-}}
\def \chplus {{\wt\chi^+}}
\def \chminus {{\wt\chi^-}}
\def \chonepm{{\wt\chi_1}^{\pm}}
\def \mchonepm{m_{\chonepm}}
\newc{\dmchi}{\Delta m_{\tilde\chi}}


\def \vlsp{\emph{VLSP}}
\def \lspj{\wt\chi_j^0}
\def \mlspj{m_{\lspj}}
\def \lspone{\wt\chi_1^0}
\def \mlspone{m_{\lspone}}
\def \lsptwo{\wt\chi_2^0}
\def \mlsptwo{m_{\lsptwo}}


\newcommand{\sell}{\tilde{\ell}}
\def \slepone{\wt\l_1}
\def \mslepone{m_{\slepone}}
\def \smuone{\wt\mu_1}
\def \msmuone{m_{\smuone}}
\def \stauone{\wt\tau_1}
\def \mstauone{m_{\stauone}}
\def \snu{\wt{\nu}}
\newcommand{\smu}{\widetilde{\mu}}
\def \msnu{m_{\snu}}
\def \msnumu{m_{\snu_{\mu}}}
\def \barsnu{\wt{\bar{\nu}}}
\def \barsnul{\barsnu_{\ell}}
\def \snul{\snu_{\ell}}
\def \mbarsnu{m_{\barsnu}}
\newc {\nuL} {\tilde{\nu}_L}
\newc {\nuR} {\tilde{\nu}_R}
\newc {\snub} {\bar{\tilde{\nu}}}
\newc {\eL} {\tilde{e}_L}
\newc {\eR} {\tilde{e}_R}
\newcommand {\qL} {\tilde{q}_L}
\newcommand {\qR} {\tilde{q}_R}
\newcommand {\uL} {\tilde{u}_L}
\newcommand {\uR} {\tilde{u}_R}
\newcommand {\dL} {\tilde{d}_L}
\newcommand {\dR} {\tilde{d}_R}
\newcommand {\cL} {\tilde{c}_L}
\newcommand {\cR} {\tilde{c}_R}
\newcommand {\sL} {\tilde{s}_L}
\newcommand {\sR} {\tilde{s}_R}
\newcommand {\tL} {\tilde{t}_L}
\newcommand {\tR} {\tilde{t}_R}
\newcommand {\stb} {\overline{\tilde{t}}_1}
\newcommand {\sbot} {\tilde{b}_1}
\newcommand {\sbotb} {\overline{\tilde{b}}_1}
\newcommand {\bL} {\tilde{b}_L}
\newcommand {\bR} {\tilde{b}_R}

\def \slepl{\wt{l}_L}
\def \mslepl{m_{\slepl}}
\def \slepr{\wt{l}_R}
\def \mslepr{m_{\slepr}}
\def \stau{\wt\tau}
\def \mstau{m_{\stau}}
\def \mul{m_{\ul}}
\def \slepton{\wt\ell}
\def \mslepton{m_{\slepton}}


\def \sql{\wt{q}_L}
\def \msql{m_{\sql}}
\def \sqr{\wt{q}_R}
\def \msqr{m_{\sqr}}
\def \sq{\wt{q}}
\newc{\squark}   {\mbox{$\wt{q}$}}
\newc{\msquark}  {\mbox{$m_{\squark}$}}
\newc{\msq}{m_{\squark}}
\newc{\sqbar}    {\mbox{$\bar{\wt{q}}$}}
\newc{\ssb}      {\mbox{$\squark\overline{\squark}$}}
\def \mul{m_{\wt{u}_L}}
\def \mur{m_{\wt{u}_R}}
\def \mdl{m_{\wt{d}_L}}
\def \mdr{m_{\wt{d}_R}}
\def \mcl{m_{\wt{c}_L}}
\def \charml{\wt{c}_L}
\def \mcr{m_{\wt{c}_R}}
\newc{\csquark}  {\mbox{$\wt{c}$}}
\newc{\csquarkl} {\mbox{$\wt{c}_L$}}
\newc{\mcsl}     {\mbox{$m(\csquarkl)$}}
\def \msl{m_{\wt{s}_L}}
\def \msr{m_{\wt{s}_R}}
\def \mbl{m_{\wt{b}_L}}
\def \mbr{m_{\wt{b}_R}}
\def \mtl{m_{\wt{t}_L}}
\def \mtr{m_{\wt{t}_R}}
\def \st{\wt{t}}
\def \lstop{\wt{t}_{1}}
\def \hstop{\wt{t}_{2}}
\def \mlstop{m_{\lstop}}
\def \mhstop{m_{\hstop}}
\def \lstoplstop{\lstop\lstop^*}
\def \hstophstop{\hstop\hstop^*}
\newc{\tsquark}  {\mbox{$\wt{t}$}}
\newc{\ttb}      {\mbox{$\tsquark\overline{\tsquark}$}}
\newc{\ttbone}   {\mbox{$\tsquark_1\overline{\tsquark}_1$}}
\newc{\mix}{\theta_{\wt t}}
\newc{\cost}{\cos{\theta_{\wt t}}}
\newc{\costloop}{\cos{\theta_{\wt t_{loop}}}}

\def \mhone{m_{H_1}}
\newc{\tb}{\tan\beta}
\newc{\vev}[1]{{\left\langle #1\right\rangle}}


\def \abot{A_{b}}
\def \atop{A_{t}}
\def \atau{A_{\tau}}
\newc{\mhalf}{m_{1/2}}
\newc{\mgut}{M_{GUT}}
\newc{\mweak}{M_{W}}
\newc{\mzero} {\mbox{$m_0$}}

\newc{\lampp}{\lam^{\prime\prime}}
\newc{\lamp}{\lam^{\prime}}
\newc{\lbp}{\lam^{\prime}}
\newc{\lbpp}{\lam^{\prime\prime}}
\newc{\lb}{\lam}
\newc{\rpv}{{\not \!\! R_p}}
\newc{\rpvm}{{\not  R_p}}
\newc{\rp}{R_{p}}
\newc{\lamotho}{\lam_{131}}
\newc{\lampotho}{\lam'_{131}}
\newc{\lamdpotho}{\lam''_{131}}
\newc{\rpmssm}{{$R_p$-MSSM}\ }
\newc{\rpvmssm}{$\rpv$-MSSM}


\newc{\pelp}{\mbox{$e^+$}}
\newc{\pelm}{\mbox{$e^-$}}
\newc{\pelpm}{\mbox{$e^{\pm}$}}
\newc{\epem}{\mbox{$e^+e^-$}}
\newc{\lplm}{\mbox{$\ell^+\ell^-$}}
\def \cc {coupling constant}
\def \branch{\emph{BR}}
\def \branche{\branch(\lstop\ra be^{+}\nu_e \lspone)\times \branch(\lstop^{*}\ra \bar{b}q\bar{q^{\prime}}\lspone)}
\def \branchmu{\branch(\lstop\ra b\mu^{+}\nu_{\mu} \lspone)\times \branch(\lstop^{*}\ra \bar{b}q\bar{q^{\prime}}\lspone)}
\def\Ecm{\ifmmode{E_{\mathrm{cm}}}\else{$E_{\mathrm{cm}}$}\fi}
\newc{\rts}{\sqrt{s}}
\newc{\mev}{~{\rm MeV}}
\newc{\tev}  {\mbox{$\;{\rm TeV}$}}
\newc{\gevc} {\mbox{$\;{\rm GeV}/c$}}
\newc{\gevcc}{\mbox{$\;{\rm GeV}/c^2$}}
\newc{\intlum}{\mbox{${ \int {\cal L} \; dt}$}}
\newc{\call}{{\cal L}}
\def \miset{\not\!\!{E_T}}
\newc{\etmiss}{/ \hskip-7pt E_T}
\def \met  {\mbox{${E\!\!\!\!/_T}$}}
\def \mispt{p{\!\!\!/}_T} 
\newc{\ptmiss}{/ \hskip-7pt p_T}
\newc{\ifb}{\mbox{${\rm fb}^{-1}$}}
\newc{\ipb}{\mbox{${\rm pb}^{-1}$}}
\newc{\chis}{\mbox{$\chi^{2}$}}
\newc{\pt}{\mbox{$p_T$}}
\newc{\et}{\mbox{$E_T$}}
\newc{\dedx}{\mbox{${\rm d}E/{\rm d}x$}}
\newc{\pb}{~{\rm pb}}
\def \ppbar{p\bar{p}}
\def \hadron{\emph{hadron}}
\def \nlc{\emph{NLC }}
\def \lhc{\emph{LHC }}
\def \cdf{\emph{CDF }}
\def \d0{\emph{D0 }}
\def \tevatron{\emph{Tevatron }}
\def \lep{\emph{LEP }}
\def \jets{\emph{jets }}
\def \jet(s){\emph{jet(s) }}
\def \eslash{\not \! E}
\def \etslash{\not \! E_T }
\def \ptslash{\not \! p_T }

\def\fonec{f_{11c}} 
\def\loopdk{\lstop \ra c \lspone}
\def\brloopdk{\branch(\loopdk)}

\newc{\mplanck}{M_{\rm P}}

\def\issue(#1,#2,#3){{\bf #1} (#3) #2 } 
\def\PRD(#1,#2,#3){Phys.\ Rev.\ D \issue(#1,#2,#3)}
\def\NPB(#1,#2,#3){Nucl.\ Phys.\ B \issue(#1,#2,#3)}
\def\JP(#1,#2,#3){J.\ Phys.\issue(#1,#2,#3)}
\def\PL(#1,#2,#3){Phys.\ Lett. \issue(#1,#2,#3)}
\def\PLB(#1,#2,#3){Phys.\ Lett.\ B  \issue(#1,#2,#3)}
\def\ZP(#1,#2,#3){Z.\ Phys. \issue(#1,#2,#3)}
\def\ZPC(#1,#2,#3){Z.\ Phys. \ C  \issue(#1,#2,#3)}
\def\PREP(#1,#2,#3){Phys.\ Rep. \issue(#1,#2,#3)}
\def\PRL(#1,#2,#3){Phys.\ Rev.\ Lett. \issue(#1,#2,#3)}
\def\MPL(#1,#2,#3){Mod.\ Phys.\ Lett. \issue(#1,#2,#3)}
\def\RMP(#1,#2,#3){Rev.\ Mod.\ Phys. \issue(#1,#2,#3)}
\def\SJNP(#1,#2,#3){Sov.\ J. \ Nucl.\ Phys. \issue(#1,#2,#3)}
\def\CPC(#1,#2,#3){Comp.\ Phys. \ Comm. \issue(#1,#2,#3)}
\def\IJMPA(#1,#2,#3){Int.\ J. \ Mod. \ Phys.\ A \issue(#1,#2,#3)}
\def\MPLA(#1,#2,#3){Mod.\ Phys.\ Lett.\ A \issue(#1,#2,#3)}
\def\PTP(#1,#2,#3){Prog.\ Theor.\ Phys. \issue(#1,#2,#3)}
\def\RMP(#1,#2,#3){Rev.\ Mod.\ Phys. \issue(#1,#2,#3)}
\def\NIMA(#1,#2,#3){Nucl.\ Instrum.\ Methods \ A \issue(#1,#2,#3)}
\def\JHEP(#1,#2,#3){J.\ High \ Energy \ Phys. \issue(#1,#2,#3)}
\def\EPJC(#1,#2,#3){Eur.\ Phys.\ J. \ C \issue(#1,#2,#3)}
\def\RPP (#1,#2,#3){Rept.\Prog.\Phys \issue(#1,#2,#3)}
\newc{\PRDR}[3]{{Phys. Rev. D} {\bf #1}, Rapid  Communications, #2 (#3)}

\subsection{
Probing Non-universal Gaugino masses: Prospects at the Tevatron }

\noindent Participants: Subhendu Chakrabarti, Amitava Datta  and N. K. Mondal 
\bigskip

Experiments at Fermilab Tevatron Run I \cite{teva} have obtained
important bounds on the chargino-neutralino sector of the Minimal Supersymmetric
extension of the Standard model using the clean trilepton signal. 
However,  the analyses used the universal
gaugino mass hypothesis at the GUT scale($M_G$) motivated by the minimal
supergravity model(mSUGRA).
On the other hand it is well known that even within the
supergravity framework, non-universal gaugino masses may naturally arise
if  non-minimal gauge kinetic functions \cite{nonuni} are allowed. Specific values of 
gaugino masses 
at $M_G$ are somewhat model dependent. The main purpose of this work is to use the data
from Tevatron Run I experiments to explore the possibility of
constraining  the chargino-neutralino sector of the MSSM
without assuming gaugino mass universality. Rather than restricting ourselves to specific
models, we shall focus our attention on the following 
generic hierarchies among the soft breaking parameters $M_2$ (the SU(2) gaugino
mass parameter), $M_1$ ( the U(1) gaugino mass parameter)  
 and the Higgsino mass parameter($\mu$) at the weak scale. Each pattern 
leads to a qualitatively different signal. We believe that this classification would
lead to  a systematic analysis of  Run II data without assuming gaugino
mass unification.

\vspace{0.5cm}
\noindent
A) If $M_1 < M_2 << \mu$, the clean trilepton signal trigerred by the decays
$\chonepm \ra l^{\pm} \nu \lspone$ and $\lsptwo \ra l^+ l^-  \lspone$ 
(l = e or $\mu$ ) is the dominant one.
Here $\chonepm, \lspone $ and $\lsptwo$ are the lighter chargino
(wino like), the lightest
neutralino (bino like), assumed to be the lightest supersymmetric particle  
(LSP), and the second lightest neutralino (wino like) 
respectively. For
$M_2 \approx 2 \times M_1$,  one regains the spectrum in the popular mSUGRA model
with radiative electroweak symmetry breaking, which usually guarantees 
relatively large  $\mu$. If $M_2 \approx \mu$ both $\chonepm$ and $\lsptwo$ have
strong higgsino components, but the trilepton signal may still be sizeable. 

\vspace{0.5cm}
\noindent
B) If  $M_1< \mu \lapp M_2$, the $\lspone$ is bino like, the $\lsptwo$ has a strong
higgsino component and the $\chonepm$ is wino like. 
 In this scenario the loop induced decay
 $\lsptwo \ra \gamma \lspone$ occurs with a large branching ratio(BR),
 spoiling the trilepton signal. The signature of $\chonepm - \lsptwo$ production is
a  $\gamma$ accompanied by standard model particles 
and large missing transverse energy.  

\vspace{0.5cm}
\noindent
C) If $M_2 < M_1 << \mu$, the $\chonepm$ and the $\lspone$ are wino like and
approximately
degenerate. Here spectrum is similar to the one predicted by the AMSB model.
Since  the chargino decays almost invsibly, special search strategies  are called 
for\cite{amsb}.

\vspace{0.5cm}
\noindent
D) If $\mu << M_1, M_2$, the $\chonepm, \lspone$ and $\lsptwo$ are approximately
degenerate
and higgsino like. As in C) special strategies for invisible/nearly invisible
particles should be employed.

In this working group report we shall focus on scenario A).
 The trilepton signal has the added advantage that it is independent of
the gluino mass $\mgl$ and hence independent of additional assumptions about
the SU(3) gaugino sector.

The important  parameters for the production of a $\chonepm$ - $\lsptwo$  
pair at the Tevatron
are   $ M_2, \mu$
, tan$\beta$ 
and the  masses of the L type  squarks belonging to the first
generation $\msql$, where q = u,d . The  squark masses in question can safely
be assumed to be degenerate,  as is guaranteed by the $SU(2)_L$ symmetry, 
barring small
calculable corrections due to SU(2) breaking D-terms. The parameter $M_1$
hardly affects the production cross section in scenario A) as will be shown below. 

It may be noted that the
bulk of the LEP constraints  on the electroweak gaugino sector arise due to
negative results from  chargino search. The chargino pair production 
 cross section at LEP is 
strongly suppressed for small  sneutrino masses. The most conservative 
limits are, therefore, obtained
for relatively light sleptons and sneutrinos. 

Although the production cross section at Tevatron is independent of 
slepton/sneutrino masses, the leptonic BR's of  $\chonepm$ and $\lsptwo$
depends on these masses. Since the BR's in question are relatively small
for heavy slepton/sneutrinos, the conservative limits correspond to such
choices.   Hence the information from Tevatron and LEP play complementary roles.

 Using the event generator PYTHIA \cite{pythia} and 
the kinematical cuts used in the CDF paper
\cite{teva}, we have simulated the trilepton signal for RUN I without
assuming gaugino mass unification. For the  purpose
of illustration we present a subset of our results in table 1. The details will
be presented else where\cite{cdm}.

\begin{table}[h]
\caption{ }
\hskip4pc\vbox{\columnwidth=26pc
\begin{tabular}{lllll}
$\mchonepm$(GeV) & $\mlspone (GeV)$ & $\sigma_{p}(pb)$ & $ BR $ &
$efficiency$ \\ \hline
77.23 & 50.85  & 5.41  & 0.0127  & 0.035   \\
76.25& 38.0  & 5.77 & 0.0128 & 0.076   \\
77.02& 31.0& 5.54 & 0.0129  & 0.081  \\
\end{tabular}
}
\label{rpv_allow}
\end{table}
For the calculations in table 1 we have set $\mu$ = - 400 GeV, tan $\beta$ = -6.0
and $\msquark = \mslepton $ = 1.5 TeV. $\mchonepm$ is approximately fixed at a
specific value using the parameter $M_2$ while the LSP mass is varied using the 
parameter $M_1$. The production cross section is denoted by $\sigma_p$ while BR
is the branching ratio of the produced pair to decay into the clean trilepton
channel.  

We have restricted ourselves to a relatively 
low value  of tan $\beta$ since large values
of this parameter lead to light $\tau$ - sleptons and the final state is
dominated by
$\tau$ leptons instead of e or $\mu$. It follows from table 1 that for a given
chargino mass the production cross section and the trilepton BR remains constant
to a very good approximation for different choices of $M_1$ (or the $\lspone$ mass). 
The efficiency of the kinematical cuts on the other hand increases with lowering
of $M_1$. Thus a lower limit on the mass of $\lspone$ as a function of the 
$\mchonepm$ is expected from the non-observation of any signal at Run I.

 This limit may have  important bearings on the viability of the LSP as the
dark matter candidate. The current lower limit on $\mlspone$ from LEP 
\cite{lepsusy} crucially hinges on the
gaugino mass unification hypothesis since it essestially originates from the 
chargino mass limit. Thus it is worthwhile to reexamine the limit without assuming 
gaugino mass unification. The indirect limit on $\mlspone$ without gaugino
mass unification is as low as 6 GeV \cite{boudjema}. It will also be interesting to
see  howfar this limit can be strengthened by data from direct searches  at Run I 
and Run II.

\section{Extra dimensions}

\subsection{Collider signals for Randall-Sundrum model (RS1) with SM gauge and
fermion fields
in the bulk}

\noindent Participants: K.~Agashe, K.~Assamagan,
J.~Forshaw and R.M.~Godbole
\bigskip
  
This work is based on the model in  
\cite{custodial} to which the reader is referred  
for further details and for references.  
  
Consider the Randall-Sundrum (RS1) model    
which is  
a compact slice of AdS$_5$,      
\begin{eqnarray}   
ds^2 & = & e^{-2k |\theta| r_c} \eta^{\mu \nu} dx_{\mu} dx_{\nu} + r_c^2 d   
\theta^2, \; - \pi \leq \theta \leq \pi,  
\label{metric}  
\end{eqnarray}   
where the extra-dimensional interval is realized as an orbifolded circle of   
radius $r_c$. The two orbifold fixed points, $\theta = 0, \pi$, correspond   
to the ``UV'' (or ``Planck) and ``IR'' (or ``TeV'') branes respectively. In   
warped spacetimes the relationship between  
5D mass scales and 4D mass   
scales (in an effective 4D description) depends on   
location in the extra dimension through the   
warp factor, $e^{-k |\theta| r_c}$. This allows large 4D mass hierarchies to   
naturally arise without large hierarchies in the defining 5D theory, whose  
mass parameters are taken to be of order the observed   
Planck scale, $M_{ Pl } \sim 10^{18}$ GeV.  
For example, the 4D massless graviton   
mode is localized near the UV brane    
while Higgs physics is taken to be   
localized on the IR brane. In the 4D effective theory one then finds
\begin{equation}
{\rm Weak ~Scale} \sim M_{ Pl } e^{-k \pi r_c} .   
\end{equation}  
A modestly large radius, i.e.,  
$k \pi r_c \sim \log \left( M_{ Pl } / \hbox{TeV} \right)  
\sim 30$, can then accommodate a TeV-size weak scale.    
Kaluza-Klein (KK) graviton resonances have   
$\sim k e^{ - k \pi r_c }$, i.e., TeV-scale masses since their wave functions   
are also localized near the IR brane.  
  
In the original RS1 model, it was assumed that the entire SM (including  
gauge and fermion fields) resides on the TeV brane. Thus, the effective UV cut-off  
for the gauge, fermion and Higgs fields,   
and hence the scale suppressing higher-dimensional operators, is  
$\sim$ TeV. However, bounds from electroweak precision data   
on this cut-off are $\sim 5-10$ TeV, whereas those from flavor changing neutral   
currents (FCNC's)  
(for example, $K - \bar{K}$ mixing)  
are $\sim 1000$ TeV. Thus, to stabilize the electroweak scale requires  
fine-tuning,   
i.e., even though RS1 explains the big hierarchy between  
Planck and electroweak scale, it has a ``little'' hierarchy problem.  
  
A solution to this problem is to move the  
SM gauge and fermion fields into the bulk.   
Let us begin with how bulk fermions enable us to evade flavor constraints.  
The localization of the wave function of the massless chiral mode  
of a $5D$ fermion  
(identified with the SM fermion)   
is controlled by the $c$-parameter.  
In the warped scenario, for  
$c>1/2$ ($c<1/2$) the zero mode is localized near the Planck (TeV) brane,  
whereas for $c = 1/2$, the wave function is {\em flat}.   
So, we choose $c > 1/2$ for light fermions so that the  
effective UV cut-off $\gg$ TeV and thus FCNC's are suppressed. Also  
this naturally results in a small $4D$ Yukawa coupling  
to the Higgs on TeV brane without any hierarchies in the  
fundamental $5D$ Yukawa couplings. Left-handed top and bottom quarks are 
close to $c = 1/2$  (but $< 1/2$) -- we can show  $c_L \sim 1/2$ is necessary 
to be consistent with $Z \rightarrow \bar{b}_L b_L$  
for KK masses $\sim$ few TeV -- whereas {\em right}-handed top quark is 
localized  near the TeV brane to get $O(1)$ top Yukawa coupling. 
Furthermore, few ($3-4$) TeV KK masses are  
consistent with electroweak data ($S$ and $T$ parameters)  
provided we enhance the electroweak gauge symmetry in the bulk  
to $SU(2)_L \times SU(2)_R \times U(1)_{B-L}$, thereby   
providing a custodial isospin symmetry   
sufficient to suppress excessive contributions to the $T$   
parameter.  
  
We can show that in such a  set-up (with bulk gauge fields) {\em high}-scale 
unification can be accommodated which is an added motivation for its 
consideration.  
  
In this project, our goal is to identify/study collider signals for this model.  
%
%
We can show that   
the Higgs couplings to   
electroweak gauge KK modes are enhanced (compared to that of {\em zero}-modes,  
i.e., SM gauge couplings)  
by $\sim \sqrt{k \pi r_c} \sim 5-6$ since the Higgs is localized on  
the TeV brane and the wave functions of the gauge KK mode are also peaked  
near the TeV brane.  
Thus,  
{\em longitudinal} $W,Z$ (eaten Higgs component) fusion into electroweak gauge KK modes  
(with masses $\sim$ few TeV)  
is enhanced. In turn,  
these KK modes have sizable decay widths to {\em longitudinal} $W/Z$'s:    
\begin{eqnarray}  
W_{long.} \; Z_{long.} \; (W_{long.} \; W_{long.}) & 
\stackrel{ g \sqrt{ k \pi r_c } }{ \longrightarrow }   
& W^{ \pm \; (n) }, Z^{ (n) }, \tilde{W}^{ \pm \; (n) }, Z^{ \prime \; (n) } \nonumber \\  
 & \stackrel{ g \sqrt{k \pi r_c} }{ \longrightarrow }   
& W_{long.} \; Z_{long.} \; (W_{long.} \; W_{long.} ) 
\end{eqnarray}  
(here the subscript $(n)$ denotes a KK mode).  
  
Note that the rise with energy of the $W_{long},Z_{long}$ cross section 
is softened by Higgs exchange, considerably below the energies of these 
resonances in  longitudinal $W/Z$ scattering. 

As per the AdS/CFT correspondence, this RS model is dual to a strongly coupled 
large-$N$  $4D$ conformal field theory (CFT)   
with $SU(3)_c \times SU(2)_L \times SU(2)_R \times U(1)_{ B - L } $  
global symmetry whose $SU(3)_c \times SU(2)_L \times U(1)_Y$ subgroup is gauged.  
A Higgs on the TeV brane corresponds to a composite of the CFT responsible 
for spontaneous breaking of $SU(2)_L \times SU(2)_R$ symmetry. That is, this 
model is dual to a particular type of a composite Higgs model. The electroweak 
gauge KK modes are techni-$\rho$'s in the dual interpretation. Thus, the 
enhanced  coupling of Higgs to electroweak gauge KK modes was  expected from 
their CFT dual interpretation as strongly  coupled composites.  
  
\begin{figure}[htbp]  
\epsfxsize=4cm  
\centerline{\epsfbox{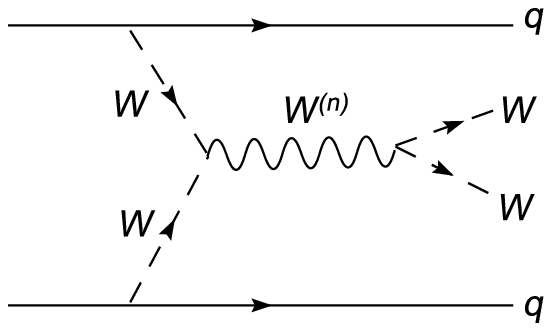},\hbox{\hskip .5cm},
\epsfbox{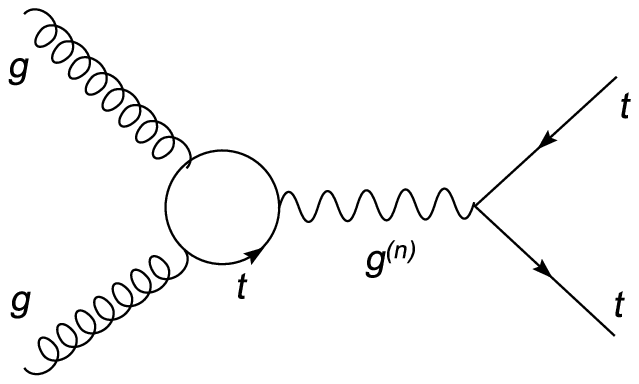}}  
\epsfxsize=4cm  
\centerline{\epsfbox{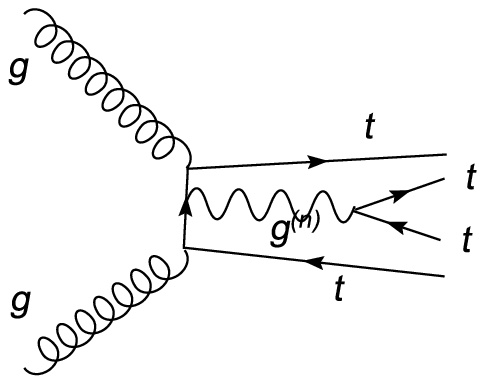},
\hbox{\hskip .5cm},  
\epsfbox{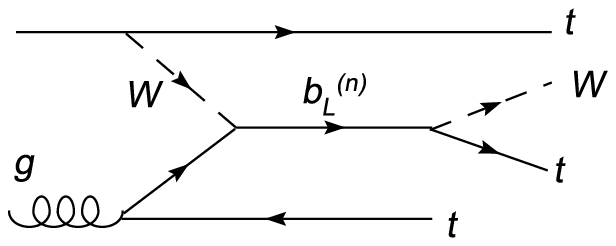}}  
\caption{Possible production mechanisms for KK states at the LHC}  
\label{fig:production}  
\end{figure}

This is similar to technicolor models where one might anticipate a   
signal at the LHC in longitudinal $W/Z$ scattering for $\sim 2$ TeV 
techni-$\rho$'s. This process is illustrated in Figure \ref{fig:production}.  
Whether there exists an observable signal for $3$ TeV gauge KK modes 
requires a calculation of the cross-section and a simulation of the process 
and associated backgrounds, which is in progress.  In particular, one needs 
to determine whether the strong coupling to these new particles can compensate 
the suppression in rate due to the largeness of the resonant mass.  
  
There are also possible signals with final states involving either  
two, three or four top quarks which are also illustrated in 
Figure \ref{fig:production}.  
All three channels benefit from the fact that $t_R$ is strongly coupled, i.e.   
$\sqrt{ k \pi r_c }$-enhanced, to the gluon and/or $W_R$   
KK modes since its wave function is localized near TeV brane. The final  
channel illustrated in Figure \ref{fig:production} benefits from an enhanced  
Higgs-$t_R$-$b^{ (n) }_L$ coupling $\sim \lambda_{ t } f (c_L)   
\sim \sqrt{10}$ (where $f (c_L)  
\approx \sqrt{ 2 / ( 1 - 2 c_L ) }$ and $c \sim 0.4$ for $(t,b)_L$) which leads to $b_L^{(n)}$  
production via longitudinal $W-t_R$ fusion. Such studies are underway.  
  
\section{Linear collider}

\subsection{\boldmath
Transverse beam polarization and CP-violating triple-gauge-boson
couplings in $e^+e^-\to \gamma Z$}
\noindent 
Participants: B. Ananthanarayan, A. Bartl, Saurabh D. Rindani, Ritesh K. Singh
\bigskip
The project was to study the benefits from significant transverse
polarization at the Linear Collider through the window of CP violation.
Two of the members of the collaboration had recently studied the
possibility of observing CP violation in the reaction $e^+ e^-\to
t\bar{t}$.  It had been concluded in that study that 
CP violation only from (pseudo-)scalar (S)
or tensor (T) type interactions due to beyond the standard model physics
could be probed in the reaction when no final state polarization
is observed, in the presence of transverse beam polarization.  
This result was obtained by generalizing certain results
due to Dass and Ross from the 1970's.  

Discussions at WHEPP8 took place around the works cited above.
It was realized that in a reaction involving self-conjugate
neutral particles in the final state, transverse beam polarization could
assist in probing CP violation that arose not necessarily from
S and T currents.  This stems from the fact that in the latter
reaction, the matrix elements for the reaction receive contribution
from the $t$ and $u$ channels.  As a result, one project that was
isolated was to carry out a full generalization of the results of
Dass and Ross that were pertinent to $s-$channel reactions, to
those which involve $t$ and $u$ channels.  

As a first step therefore, one wished to study specific examples.
For instance, the members of the collaboration wished to 
study the reaction  $e^+e^-\to \gamma Z$ as an example.  In particular,
all beyond the standard model physics was assumed to arise from
anomalous triple-gauge-boson couplings.  The task was to compute the
differential cross-section for the process in the presence of anomalous
couplings and transverse beam polarization, and then to construct
suitable CP-odd asymmetries.  A numerical study was proposed to place
suitable confidence limits on the anomalous couplings for realistic
polarization and integrated luminosity at a design LC energy of $\sqrt{s}
=500$ GeV.  

After WHEPP8 the members of the collaboration carried out the project
and the results are published in \cite{ba_plb}.
Two of the members of the
collaboration have also considered more recently the most general gauge-invariant and chirality-conserving interactions that would contribute to CP violation in $e^+e^-\to \gamma Z$ \cite{contact}.

Another possible example that was considered by the members of the
collaboration was a reaction with a slepton pair in the final state.
Work is yet to begin on this.

\subsection{\boldmath
Decay lepton angular distribution in top production -- \\ decoupling from
anomalous $tbW$ vertex}
\bigskip
\noindent Participants:
 Rohini M. Godbole, Manas Maity, Saurabh D. Rindani, Ritesh K. Singh
\bigskip

The project was to study the (in)dependence of decay lepton angular distribution,
on any anomalous coupling in top-decay vertex, for different production 
processes of the top-quark. It is known in literature that the
angular distribution of decay lepton, in pair production of top-quarks, is
independent of the anomalous $tbW$ coupling to linear order. This result is 
independent of the initial state and hence valid for all colliders. Thus
decay lepton angular distribution provides, at all colliders, a pure probe of 
possible anomalous interaction in the pair production of top-quarks, 
uncontaminated by any new physics in decay of top-quark. This result, though 
very attractive and useful, lacks a fundamental understanding. At WHEPP-8, we 
discussed possible approaches to understand the above said decoupling and 
explore the possibilities of extending this ``decoupling theorem" to processes 
involving single top production and top pair production in $2\to3$ processes.
If the decoupling is observed in $2\to3$, it possibly can be 
extended to $2\to n$ processes of top production.

\subsection{\boldmath Graviton Resonances in ${ e^+e^- \to \mu^+\mu^-}$
with\\ beamstrahlung and ISR}

\noindent Participants: Rohini M. Godbole, Santosh Kumar Rai and 
Sreerup Raychaudhuri 
\bigskip

The next generation of high-energy $e^+ e^-$ colliders~\cite{tesla,others}
will necessarily be Linear Colliders to avoid losses due to synchrotron 
radiation. 
However, as a linear collider will have single-pass colliding beams, 
the bunches constituting a beam would have to be focused to very small
dimensions to get an adequate luminosity. This is an essential part
of the design of all the proposed machines. The high density of charged
particles at the interaction point would necessarily be accompanied by 
strong electromagnetic fields.  The interactions of beam particles with 
the accelerating field generates the so-called {\it initial state radiation} 
(ISR), while their interactions with the fields generated by the other beam 
also generates radiation, usually dubbed {\it Beamstrahlung}\cite{beamold}.

Traditionally, ISR and Beamstrahlung have been considered nuisances which 
cause energy loss and disrupt the beam collimation. The energy-spread due 
to these radiative effects has led to a requirement of realistic simulations 
for physics processes which would require the knowledge of the energy spectrum 
of the colliding beams at the interaction point. The beam designs being
considered are usually such that these effects are minimized.

In this note we argue that instead of just being a nuisance which we have
to live with, photon radiation from initial states can actually be of great use
in probing new physics scenarios under certain circumstances.
As a matter of fact tagging with ISR photons has been used effectively in the 
LEP experiments, to search for final states which do not leave too much visible 
energy in the detectors; for example, a  $\tilde \chi^+ \tilde \chi^-$ 
(chargino) pair with $\tilde\chi^+ $ and the LSP $\tilde \chi^0_1$  being 
almost degenerate\cite{Chen:1995yu}. Here, we look at a different aspect and 
usage of these radiative effects. To illustrate it we look at one of the 
simplest processes at an $e^+ e^-$ collider, viz.
$$    
e^+ ~e^- \to X^* \to \mu^+ ~\mu^- 
$$
where $X$ can be either a massive scalar, vector or tensor. In the Standard 
Model, $X = \gamma, Z$. For any heavy particle $X$, there will be resonances
in the $s$-channel process, observable as peaks in the invariant mass
$M_{\mu^+\mu^-}$ distribution. At LEP, for example, this process was used
to measure the $Z$-resonance line shape. In this note, we focus our 
attention on tensor particle resonances, the tensors being the massive 
Kaluza-Klein gravitons as predicted in the well-known braneworld model of
Randall and Sundrum\cite{RS}. 

The central point in our argument is that it is very likely that the
next generation linear colliders would run at one (or a few) fixed
value(s) of
centre-of-mass energy. For example, Tesla\cite{tesla} is planned to run at 
$\sqrt{s} = 500$~GeV and 800~GeV. However, the predicted massive graviton
excitations of the Randall-Sundrum (RS) model may not lie very close to
these energy values. Consequently, the new physics effect due to exchange
of RS gravitons will be off-resonance and hence strongly suppressed. 
However, a spread in beam-energy would cause {\it some} of the events 
to take place at an effective (lower) centre-of-mass energy around the 
resonance(s) and hence provide an enhancement in the cross-section. 
A similar effect, for example, was observed in $Z$-resonances at LEP-1.5 
and dubbed the `return to the $Z$-peak'. We, therefore, investigate 
`return to the {\it graviton} peak' in the process 
$e^+ ~e^- \to \mu^+ ~\mu^-$.

\begin{figure}[htbp]
\epsfxsize=6cm
\centerline {\epsfbox{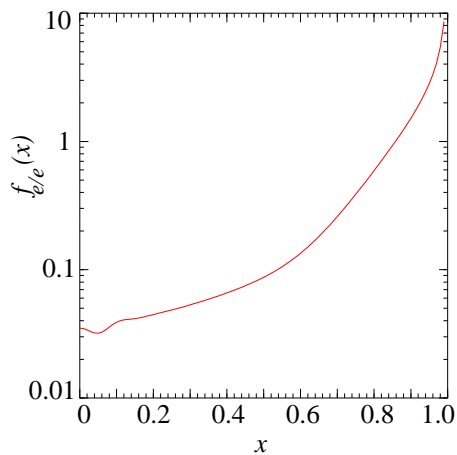}}
\caption{Illustrating the effective electron luminosity at Tesla-800
as a function of $x = E_e/E_{\rm beam}$, the energy fraction of the
electron after radiating a photon. The Beamstrahlung parameter is 
$\Upsilon = 0.09$.}
\label{fig:Fig1}
\end{figure}

In our analysis of radiative effects we use the structure function formalism
for ISR and Beamstrahlung developed in Refs.\cite{chen,rohini}. Specifically,
we use the expression for the electron spectrum function 
presented in \cite{rohini}. 
Figure \ref{fig:Fig1} shows the electron energy spectrum for the given design
parameters for the linear collider at TESLA \cite{tesla} running at 
$\sqrt{s} = 800$~GeV. It is worth noting that the large spread in the
distribution function is more due to Beamstrahlung than to ISR
effects.\cite{rohini}

In the two-brane model of Randall and Sundrum, the Standard Model is augmented
by a set of Kaluza-Klein excitations of the graviton, which behave like massive 
spin-2 fields with masses $M_n = x_n m_0$, where the $x_n$ are the zeroes of the
Bessel function of order unity, $n$ is a non-negative integer and $m_0$ is an 
unknown mass scale close to the electroweak scale. Search possibilities for
these gravitons at future $e^+e^-$ colliders, have been studied in the 
literature \cite{SKR}. Experimental data from the Drell-Yan process at the
Tevatron constrain $m_0$ to be more-or-less above 130~GeV\cite{tevatron}. 
Another
undetermined parameter of the theory is the curvature of the fifth dimension, 
expressed as a fraction of the Planck mass $c_0 = {\cal K}/M_P$. Feynman rules
for the Randall-Sundrum graviton excitations can then be read-off from the 
well-known Feynman rules given in Ref.\cite{Han} by making the simple 
substitution $\kappa \to 4\sqrt{2}\pi c_0/m_0$. Noting that the massive 
graviton states exchanged in the $s$-channel can lead to Breit-Wigner 
resonances, it is now a straightforward matter to calculate the cross-section 
for the process $e^+ ~e^- \to \mu^+ ~\mu^-$ and implement it in a simple Monte 
Carlo event generator. 

\begin{figure}[htb]
\epsfxsize=13cm
\centerline{\epsfbox{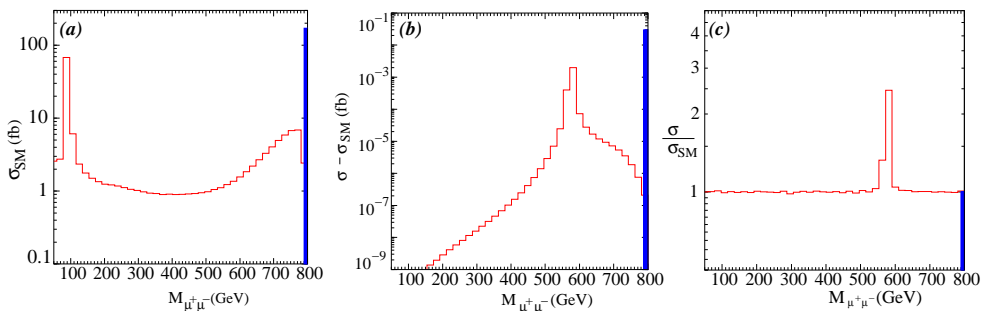}}

\caption{ Invariant mass distribution with (outlined histogram) and 
without (filled histogram)
radiative effects.
The figures show
(a)~the bin-wise cross-section $\sigma_{SM}$ for the SM background,~
(b)~the bin-wise excess cross-section $\sigma - \sigma_{SM}$ predicted
in the RS model and~
(c)~the signal-to-background ratio $\sigma/\sigma_{SM}$.}
\label{fig:Fig2}
\end{figure}

Our numerical analysis has 
been performed for values $m_0 = 150$~GeV and $c_0 = 0.01$, which
implies that the lightest ($n = 1$) massive excitation has mass 
$M_1 = 574.5$~GeV, putting it well beyond the present reach of 
Run-II data at the Tevatron\cite{tevatron}. With this choice, however, the next
excitation is predicted to have mass $M_2 > 1$~TeV, which puts it well 
beyond the reach of Tesla-800. We expect, therefore, to detect one, and 
only one, resonance.  The value of $c_0$ has been chosen at the lower end 
of the possible range, since this leads to a longer lifetime
for the Kaluza-Klein state and hence a sharper resonance in the cross-section. 
Following standard practice for linear collider studies, we eliminate most of
the backgrounds from beam-beam interactions and two-photon processes by imposing
an angular cut $10^0 < \theta_{\mu^\pm} < 170^0$ on the final-state muons.
Some of our results are illustrated in Figure \ref{fig:Fig2}, which shows the 
binwise 
distribution of invariant mass $M_{\mu^+\mu^-}$ of the (observable) final state.
In Figure~2($a$), we have plotted the distribution predicted in the Standard 
Model (SM). Figure~2($b$) shows the excess over the SM prediction expected
in the Randall-Sundrum model and Figure~2($c$) shows the signal-to-background 
ratio.

At a linear collider with fixed center-of-mass energies, all the events
for the above process should be concentrated in a single invariant mass bin at
$M_{\mu^+\mu^-} \sim \sqrt{s}$ in the lab frame. In Figure~2 these correspond to
the solid (blue) bins in the $M_{\mu^+\mu^-}$ distribution. Comparison of 
Figures 2($a$) and 2($b$) show that the expected signal is very small indeed, 
about 1 in $10^4$.  Consequently the ratio exhibited in Figure 2($c$) is almost 
precisely unity. This is because our parameter choice leads to a graviton of 
mass 574.5~GeV, and decay width of a few GeV, which is far away from the 
centre-of-mass energy $\sqrt{s} = 800$~GeV.

The outlined (red) histograms in Figure~2 show the invariant mass 
distribution when 
radiative effects are included. It is immediately apparent that the invariant 
mass and hence the effective centre-of-mass energy 
$\sqrt{s'} = M_{\mu^+\mu^-}$ is spread out from
the beam energy $\sqrt{s}$. In ($a$), we can see a distinct peak at the lower 
end which represents `return to the $Z$-peak'. The cross-section for this peak 
is not as high as one might expect for a narrow resonance like the $Z$ because 
this corresponds to an extremely large value for the energy fraction $x$ taken 
away by the photon, for which the luminosity is extremely small. 
The shape of the rest of the histogram is simply a reflection of the electron 
luminosity shown in Figure~1. A similar phenomenon happens in ($b$) 
due to the large 
spread in the energy of the colliding beams. Here the radiative 
return to the resonant KK-graviton is quite apparent. In fact, excitation 
of the graviton resonance leads to a greatly enhanced cross-section, as this 
graph shows. The outlined (red) histogram in ($c$), shows the 
signal-to-background ratio. 
This ratio removes the $Z$-peak and clearly throws into prominence the 
graviton resonance, presenting us with a clear signal for a new resonant 
particle. To confirm that it is indeed a graviton, one must run various tests, 
such as plotting the angular distribution. These will be discussed in a 
forthcoming publication\cite{GRR}. Note also that the method can be used with
effect only for final states not involving strongly interacting particles, as
 two-photon processes can give rise to a substantial two-jet production for
invariant masses quite a bit smaller than the nominal centre of mass 
energy of the collider~\cite{rohini}.

It is thus clear that ISR and Beamstrahlung can play a non-trivial role in the
identification of new physics effects. This is a positive feature of these 
radiative phenomena, which has not often been considered, and the main purpose 
of this work is to emphasize this aspect.

\def\cO#1{{\cal{O}}\left(#1\right)}
\newcommand{\be}{\begin{equation}}
\newcommand{\ee}{\end{equation}}
\newcommand{\br}{\begin{eqnarray}}
\newcommand{\er}{\end{eqnarray}}
\newcommand{\ba}{\begin{array}}
\newcommand{\ea}{\end{array}}
\newcommand{\bi}{\begin{itemize}}
\newcommand{\ei}{\end{itemize}}
\newcommand{\bn}{\begin{enumerate}}
\newcommand{\en}{\end{enumerate}}
\newcommand{\bc}{\begin{center}}
\newcommand{\ec}{\end{center}}
\newcommand{\ul}{\underline}
\newcommand{\ol}{\overline}
\newcommand{\sm}{${\cal {SM}}$}
\newcommand{\as}{\alpha_s}
\newcommand{\aem}{\alpha_{em}}
\newcommand{\ycut}{y_{\mathrm{cut}}}
\newcommand{\susy}{{{SUSY}}}
\newcommand{\Dir}{\kern -6.4pt\Big{/}}
\newcommand{\Dirin}{\kern -10.4pt\Big{/}\kern 4.4pt}
\newcommand{\DDir}{\kern -10.6pt\Big{/}}
\newcommand{\DGir}{\kern -6.0pt\Big{/}}
\def\Ecm{\ifmmode{E_{\mathrm{cm}}}\else{$E_{\mathrm{cm}}$}\fi}
\def\gluino{\ifmmode{\mathaccent"7E g}\else{$\mathaccent"7E g$}\fi}
\def\photino{\ifmmode{\mathaccent"7E \gamma}\else{$\mathaccent"7E \gamma$}\fi}
\def\mgluino{\ifmmode{m_{\mathaccent"7E g}}
             \else{$m_{\mathaccent"7E g}$}\fi}
\def\taugluino{\ifmmode{\tau_{\mathaccent"7E g}}
             \else{$\tau_{\mathaccent"7E g}$}\fi}
\def\mphotino{\ifmmode{m_{\mathaccent"7E \gamma}}
             \else{$m_{\mathaccent"7E \gamma}$}\fi}
\def\ML{\ifmmode{{\mathaccent"7E M}_L}
             \else{${\mathaccent"7E M}_L$}\fi}
\def\MR{\ifmmode{{\mathaccent"7E M}_R}
             \else{${\mathaccent"7E M}_R$}\fi}
\def\lsim{\buildrel{\scriptscriptstyle <}\over{\scriptscriptstyle\sim}}
\def\gsim{\buildrel{\scriptscriptstyle >}\over{\scriptscriptstyle\sim}}
\def\issue(#1,#2,#3){{\bf #1}, #2 (#3)} 
\def\jp #1 #2 #3 {{J.~Phys.} {#1} (#2) #3}
\def\pl #1 #2 #3 {{Phys.~Lett.} {#1} (#2) #3}
\def\np #1 #2 #3 {{Nucl.~Phys.} {#1} (#2) #3}
\def\zp #1 #2 #3 {{Z.~Phys.} {#1} (#2) #3}
\def\pr #1 #2 #3 {{Phys.~Rev.} {#1} (#2) #3}
\def\prep #1 #2 #3 {{Phys.~Rep.} {#1} (#2) #3}
\def\prl #1 #2 #3 {{Phys.~Rev.~Lett.} {#1} (#2) #3}
\def\mpl #1 #2 #3 {{Mod.~Phys.~Lett.} {#1} (#2) #3}
\def\rmp #1 #2 #3 {{Rev. Mod. Phys.} {#1} (#2) #3}
\def\sjnp #1 #2 #3 {{Sov. J. Nucl. Phys.} {#1} (#2) #3}
\def\cpc #1 #2 #3 {{Comp. Phys. Comm.} {#1} (#2) #3}
\def\xx #1 #2 #3 {{#1}, (#2) #3}
\def\AJ(#1,#2,#3){Astrophysical.\ Jour. \issue(#1,#2,#3)}
\def\NPB(#1,#2,#3){Nucl.\ Phys.\ B \issue(#1,#2,#3)}
\def\NPPS(#1,#2,#3){Nucl.\ Phys.\ Proc. \ Suppl \issue(#1,#2,#3)}
\def\NP(#1,#2,#3){Nucl.\ Phys.\ \issue(#1,#2,#3)}
\def\PL(#1,#2,#3){Phys.\ Lett.\ \issue(#1,#2,#3)}
\def\PLB(#1,#2,#3){Phys.\ Lett. \ B \issue(#1,#2,#3)}
\def\PRL(#1,#2,#3){Phys.\ Rev.\ Lett. \issue(#1,#2,#3)}
\def\PRD(#1,#2,#3){Phys.\ Rev.\ D \issue(#1,#2,#3)}
\def\JHEP(#1,#2,#3){J.\ High \ Energy \ Phys. \issue(#1,#2,#3)}
\def\preprint{{preprint}}
\def\Ord{\lower .7ex\hbox{$\;\stackrel{\textstyle <}{\sim}\;$}}
\def\OOrd{\lower .7ex\hbox{$\;\stackrel{\textstyle >}{\sim}\;$}}
\def\MCH {$\tilde\chi_1^+$}
\def \CH{{\tilde\chi}^{\pm}}
\def \lsp{\tilde\chi_1^0}
\def \SNU{\tilde{\nu}}
\def \BARSNU{\tilde{\bar{\nu}}}
\def \MLSP{m_{{\tilde\chi_1}^0}}
\def \MCH{m_{{\tilde\chi}^{\pm}}}
\def \MCHMIN {\MCH^{min}}
\def \ET{\not\!\!{E_T}}
\def \LL{\tilde{l}_L}
\def \LR{\tilde{l}_R}
\def \MLL{m_{\tilde{l}_L}}
\def \MLR{m_{\tilde{l}_R}}
\def \MSNU{m_{\tilde{\nu}}}
\def \PROCESS{e^+e^- \rightarrow \tilde{\chi}^+ \tilde{\chi}^- \gamma}
\def \PI{{\pi^{\pm}}}
\def \DM{{\Delta{m}}}
\newcommand{\bQ}{\overline{Q}}
\newcommand{\ad}{\dot{\alpha }}
\newcommand{\bd}{\dot{\beta }}
\newcommand{\dd}{\dot{\delta }}
\def \CH{{\tilde\chi}^{\pm}}
\def \MCH{m_{{\tilde\chi}_1^{\pm}}}
\def \LSP2{\tilde\chi_2^0}
\def \MUL{m_{\tilde{u}_L}}
\def \MUR{m_{\tilde{u}_R}}
\def \MDL{m_{\tilde{d}_L}}
\def \MDR{m_{\tilde{d}_R}}
\def \MSNU{m_{\tilde{\nu}}}
\def \MLL{m_{\tilde{l}_L}}
\def \MLR{m_{\tilde{l}_R}}
\def \MQL{m_{\tilde{q}_L}}
\def \MQR{m_{\tilde{q}_R}}
\def \mhf{m_{1/2}}
\def \MST{m_{\tilde t_1}}
\def \RPVC{\lambda'}
\def\tth{\tilde{t}\tilde{t}h}
\def\qqh{\tilde{q}_i \tilde{q}_i h}
\def\t1{\tilde t_1}
\def \pt{p{\!\!\!/}_T}  
\newcommand{\etal}{{\it et al.}\ }
\def\lapp{\mathrel{\rlap{\raise.5ex\hbox{$<$}}
                    {\lower.5ex\hbox{$\sim$}}}}
\def\gapp{\mathrel{\rlap{\raise.5ex\hbox{$>$}}
                    {\lower.5ex\hbox{$\sim$}}}}

\subsection{
Probing R-parity Violating Models of Neutrino
Mass at the Linear Collider
}

\noindent Participants: A.Bartl, S. P. Das, A. Datta, R. M. Godbole and D. P. Roy 
\bigskip

The observation  of neutrino oscillations and the measurement 
of oscillation parameters by the SUPERK collaboration
\cite{SK} and  others\cite{other,dp} have  established that at 
least two of the  neutrino  masses are non-zero
albeit their magnitudes  are several orders of magnitude 
smaller than that of the other fermions. 

A natural explanation of the
smallness of the neutrino masses is perhaps the most challenging
task of current high energy physics. The see-saw mechanism 
\cite{seesaw} is certainly  the most popular model. However,
the simplest version of this model -  a supersymmetric grand unified
theory (SUSYGUT) of the grand desert type, which can also explain
 coupling
constant unification\cite{amaldi}, has practically no 
other crucial prediction for  TeV scale
physics. If an intermediate scale is allowed then both SUSY and
non-SUSY GUTs, the latter being plagued by the notorious fine tuning
problem, may  serve the purpose. But there is no 
 compelling reason within the framework of these models 
either for new physics at the TeV scale. 

In contrast within the framework of   R-parity violating (RPV) SUSY
Majorana masses of the neutrinos can be generated both at the tree
level and at the one loop level quite naturally. More importantly
the physics of this mechanism is entirely governed by TeV scale physics
(sparticle masses and couplings) which can in principle be verified at
the next round of collider experiments.

Neutrino masses within the framework of RPV SUSY have been studied
by several groups\cite{numass}. Such masses may arise both at the tree
level as well as at the one loop level. As an example, we refer
to \cite{abada}, where upper bounds on RPV bilinear and trilinear terms 
were derived ( see tables III - VIII of \cite{abada} )using some 
simplifying assumptions about the R-parity-conserving (RPC) 
sector(see below).

In this working group project  we try to  further sharpen these predictions. 
We obtain  several combinations of lepton number violating trilinear 
($\lambda_{ijk}, \lambda^{\prime}_{ijk} $, i,j,k = 1,2,3) 
and bilinear ($\mu_i$, i=1,2,3) couplings which are consistent with  
the current ranges of the oscillation parameters \cite{altarelli}.

The squared mass differences of different neutrinos  are defined as:
\be
\Delta m^2_{sun}\equiv \vert\Delta m^2_{12}\vert  ,~~~~~~~
\Delta m^2_{atm}\equiv \vert\Delta m^2_{23}\vert~~~.
\label{eq:fre}
\ee   where $\Delta m^2_{12}= m_2^2- m_1^2 > 0$ 
and $\Delta m^2_{23}= m_3^2- m_2 ^2$ assuming 
$m_1^2 < m_2^2 < m_3^2$. 

The limits on them  are 
\[
5\times 10^{-5}  <\Delta m^2_{sun} (eV^2)<  10\times 10^{-5}\]
  and 
\[ 1\times 10^{-3}  <\Delta m^2_{atm}(eV^2) < 4\times 10^{-3}. \]
Similarly the mixing angle
constraints are 
$$0.29<\tan^2\theta_{12}<0.82,$$
$$0.45<\tan^2\theta_{23}<2.3,$$
$$0.0<\tan^2\theta_{13}< 0.05,$$ 
for solar, atmospheric and CHOOZ data
\cite{altarelli} .

Since our results are basically illustrative, we employed the  
same  simplifying assumptions as in \cite{abada}. 
\begin{itemize}
\item  All masses and mass parameters in the RPC sector of the MSSM
are $\approx$ 100 GeV. 
\item tan $\beta$ = 2
\end{itemize} 
This leads to the following tree level and loop level mass matrices 
\cite{abada}:
\br\hspace*{-0.5cm}
{\cal{M}}^{\mathrm{tree}}_{\nu_{ij}}= C  \mu_i  \mu_j ,
 \label{MRptree}
\er 
 \br\hspace*{-0.5cm}
 {\cal{M}}^{\mathrm{loop}}_{\nu}= \!\!\left(
  \begin{array}{lll}
   K_1\lambda^2_{133}+K_2\lambda'^2_{133}&
   K_1\lambda_{133}\lambda_{233}+
   K_2\lambda'_{133}\lambda'_{233} &
    K_2\lambda'_{133}\lambda'_{333}\\
     K_1\lambda_{133}\lambda_{233}+
     K_2\lambda'_{133}\lambda'_{233}&
      K_1\lambda^2_{233}+K_2\lambda'^2_{233}&
       K_2\lambda'_{233}\lambda'_{333} \\
       K_2\lambda'_{133}\lambda'_{333} &
       K_2\lambda'_{233}\lambda'_{333}
       & K_2\lambda'^2_{333}
        \end{array}
	 \right),
	 \label{MRploop}
	  \er
where the constants are given by C $\approx$ 5.3 $\times$ $10^{-3}$
$GeV^{-1}$, $K_1$ $\approx$ 1.8 $\times$ $10^{-4}$
$GeV$ and $K_2$ $\approx$ 4.7 $\times$ $10^{-3}$
$GeV$. In \cite{abada} several scenarios were considered with five 
non-zero RPV couplings (see table III of \cite{abada}). For the purpose of
illustration we consider scenario 1 
where the  non-vanishing parameters are the three $\mu$'s,  
$\lambda_{133}$ and $\lambda_{233}$.

Now we try to fit the  above oscillation parameters by 
varying the above five parameters randomly
subject to the existing bounds ( see table V of \cite{abada};
we have considered the MSW large mixing angle solution only). 
By generating 10000  sets of parameters
we have obtained  only 3  solutions  in ( see, table \ref{rpv_allow2} ).
It is gratifying  to note that even the rather loose 
constraints on the oscillation parameters currently available 
are sufficiently restrictive to yield a remarkably  small set of solutions.

\begin{table}[h]
\caption{ 
Allowed RPV parameters consistent with the  Neutrino data in 
\protect\cite{altarelli} }
\hskip4pc\vbox{\columnwidth=26pc
\begin{tabular}{llllll}
Solution No. & $\mu_1 (GeV)$ & $\mu_2 (GeV)$ & $\mu_3 (GeV)$ & 
$\lambda_{133}$ & $\lambda_{233}$\\ \hline
I &  $1.1E-05$ & $5.9E-05$  & $8.2E-05$  &  $1.5E-04$  & $1.9E-04$ \\
II & $4.4E-06$ & $6.5E-05$ &  $8.0E-05$ &  $1.4E-04$  & $2.1E-04$ \\
III& $8.0E-06$ & $4.3E-05$ &  $7.9E-05$  &  $1.6E-04$  & $2.1E-04$ \\
\end{tabular}
}
\label{rpv_allow2}
\end{table}

Although $\lsp$ (LSP) decay is generic in RPV models, the above examples 
illustrate that the branching ratios  and the life time  of the LSP, 
which we assume to be the lightest neutralino,  
will have very specific patterns if the oscillation constraints are
imposed. In the scenario under consideration the allowed decay modes are:
\be 
(a) \lsp \ra e \tau \ET  , (b) \lsp \ra \mu \tau \ET  
~~and~~ (c) \lsp \ra \tau \tau \ET ,
\ee

where the missing energy ($\ET$) is carried by the neutrinos. 
Charge conjugate modes are included in our analysis. 

In Table \ref{dklength} we have presented some
LSP decay characteristics  calculated  by CompHEP\cite{comp} 
using the first two solutions in Table \ref{rpv_allow2}. We find that 
in order to distinguish solutions number I from II , the BR(a),(b) and 
the decay length have to be measured with accuracy better than 
17.4\%, 8.5\% and  5.6\% respectively.  

\begin{table}[h]
\caption{
Lightest Neutralino decay  branching ratios 
and decay lengths for the first two scenario  of Table \ref{rpv_allow2}.
}
\hskip4pc\vbox{\columnwidth=26pc
\begin{tabular}{llll}
Solution. No. & BR &Decay length ( c $\times  \tau$ in cm )   
\\ 
\hline
I & (a) 0.186  &   \\
  & (b) 0.323   & 35.82 \\
  & (c) 0.491   & 
\\
\hline
II & (a)0.156   &  \\
  & (b)0.352      & 37.66\\
  & (c)0.492     &
\\
\hline

\end{tabular}
}
\label{dklength}
\end{table}

Although our calculations were based on very specific assumptions
there are reasons to believe  that the restricted nature of the 
predictions of this model will continue to hold even without these 
assumptions. Improvement in the precision of the magnitudes of the 
oscillation parameters in the future long base line experiments
will impose even tighter constraints on model parameters. For example,
we have tested that the ranges of oscillation parameters in \cite{abada}
(see Table I) based on old data lead to many more solutions. 

Moreover  measurements  of superparticle masses, couplings and some of the
Branching Ratios (BRs) will be available~\cite{atlastdr} from LHC within the 
first  few years of its running, if SUSY exists. This may enable one to fix 
the constants C, $K_1$ and $K_2$ within reasonable ranges without additional
assumptions. Precision measurements  of  LSP decay properties  to 
verify the RPV models of neutrino mass seem to be a challenging, but perhaps 
feasible, task for the next linear collider (LC). 

It is interesting to analyze the 
possible information one may seek at the Linear Collider to be able to do 
this job. In case it is the RPV version of SUSY that is realized in nature, 
even the LHC will offer a rather good measurement of the mass of the LSP, 
particularly if the $\lambda, \lambda^{'}$ RPV couplings  are the dominant
ones. The LC will on the other hand will  offer a chance for  accurate 
measurement of the LSP  mass as well as its life time provided the RPV 
couplings are large and the LSP has macroscopic decay length. We see from
Table~II,  that for the particular solutions path lengths of a few cms. are 
possible for the LSP. Studies of possible accuracies of such measurement
need to be performed. It will be possible to measure the mass of the decaying 
LSP  at an LC  using either the kinematic end point measurements and/or 
through the threshold scans. Very preliminary studies~\cite{ggr} of the 
possibilities of the mass measurements of the LSP in the production of 
$\tilde \chi_1^0 \tilde \chi_2^0, \tilde \chi_1^+ \tilde \chi_1^-$, followed 
by the decay of the $\tilde \chi_2^0, \tilde \chi_1^\pm$ and the LSP in the
end exist. These studies need to be refined. Further, the LSP decay may 
also depend  on the masses of the third generation sparticles and mixing, 
precision information for which may also be available only from the LC.
These features as well as the possible interplay between the LHC and the LC 
to pin down RPV SUSY as the origin of neutrino mass need to be studied.
Finally we note that if the 
$\lambda^{'}$ couplings are indeed ${\cal O}(10^{-4})$ as required by models 
of $\nu$ mass \cite{abada}, lighter top squark decays may provide 
additional evidence in favour of these models \cite{admgspd}

\end{document}